\documentclass[preprint,12pt]{elsarticle}
\usepackage[margin=2.5cm]{geometry}
\usepackage{lipsum}
\usepackage{hyperref}
\usepackage{times}
\usepackage{amsmath}
\usepackage{amsmath,bm}
\usepackage{amssymb}
\usepackage{setspace}
\usepackage{graphics}
\usepackage{fancyhdr}
\usepackage{rotating}
\usepackage{color}
\usepackage{amsmath}
\usepackage{titlesec}
\titleformat{\paragraph}
{\normalfont\normalsize}{\theparagraph}{1em}{}
\titlespacing*{\paragraph}
{0pt}{3.25ex plus 1ex minus .2ex}{1.5ex plus .2ex}
\titleformat*{\subsubsection}{\normalfont}
\usepackage{multirow}
\usepackage{placeins}
\usepackage[table,xcdraw]{xcolor}

\journal{arXiv and prepared using Elsevier’s {\LaTeX} package}

\begin{document}
\begin{frontmatter}

\title{Vortex-Stretching based Large Eddy Simulation Framework for Wind Farms}

\author[1]{Jagdeep Singh \corref{cor1}}%
\ead{jagdeeps@mun.ca}
\author[2]{Jahrul Alam }
\ead{alamj@mun.ca}

\address{Department of Mathematics and statistics \\ Memorial University of Newfoundland, St. John's, Canada- A1C 5S7}

\cortext[cor1]{Corresponding Author.}
\begin{abstract}
In large wind farms, wake distribution behind a wind turbine causes a considerable reduction of wind velocity for downstream wind turbines, resulting in a significant amount of power loss. Therefore, it is very crucial to predict wind turbine wakes efficiently. Thus, we propose a large-eddy simulation (LES) methodology, which takes the vorticity stretching to model transients in wind turbine wakes. In addition, we present an improved actuator disk model, which accounts for two-way feedback between the atmosphere and the wakes. First, we show that the vertical profile of the mean wind predicted with the new model has an excellent agreement with experimental measurements. Next, we validate the predicted Reynolds stresses against wind tunnel data and show that the dispersive stresses account for about 40\% of Reynolds stresses. Finally, we show that the proposed LES method accurately predicts the characteristics of wind turbine wakes. Comparing the LES results with previously reported data, we have found that the new LES framework accurately predicts the flow statistics in both the near-wake and the far-wake regions.
\end{abstract}

\end{frontmatter}
\section{INTRODUCTION}
Large wind farms are increasing worldwide to mitigate climate change and accomplish sustainable developments \cite{porte2020}. However, the movement of air currents through wind farms is not well understood~\cite{miller2018,meneveau2019}. Large-eddy simulations (LES) of wind farms provide critical data that could help wind energy developers increase energy production~\cite{porte2013,churchfield2012,xie2017}. The development of technologies for reducing the power losses caused by wind turbine wake is one of the most challenging topics~\cite{Archer2018}. LES techniques can simulate details of the tip vortices and wind turbine blades. However, computational resources are insufficient for such a highly resolved LES of wind farms~\cite{stevens2017}. Predictions of near turbine regions with state-of-the-art LES methods differ significantly~\cite{stevens2018}. Flattened near wake profiles and sharp deviation from the ground surface are typically seen in the classical actuator disk model (ADM), where the local variation of the thrust across the rotor is neglected. Such errors are adequately controlled with an actuator line model if the nacelle, rotational effects of turbines, tip vortices, and the atmospheric boundary layer (ABL) region below the rotor bottom are sufficiently resolved~\cite{porte2011}. However, neither of such methods is appropriate for LES of utility-scale large wind farms in which two-way feedback of atmospheric boundary layer becomes essential~\cite{churchfield2012,xie2017,alam2019}. Therefore, a scale-adaptive LES using an improved actuator disk model can be a promising solution to balance the efficiency and accuracy. In this work, we evaluate and demonstrate the predictive capabilities of a scale-adaptive LES model for simulating wind farms using wind tunnel measurements. 

Wind farms extract energy from the lower part of the ABL, where turbulence is highly anisotropic due to different dynamical processes occurring in that region~\cite{garratt1994atmospheric,stoll2020large}. One such process is vorticity production due to the stretching of fluid elements~\cite{davidson2015turbulence,bhuiyan2020}. Thus, this article aims to present a scale-adaptive turbulence model for the LES of wind farms. In LES, one seeks to model the residual stress $( \bm{\tau})$ and, consequently, the energy flux $(\bm{\Pi})$. These two quantities ($\bm\tau$ and $\Pi$) account for the errors associated with the missing flow physics (due to limitations in measurements or LES) in the resolved (or measured) quantities. Vortex stretching and strain self-amplification are the two most critical dynamical mechanisms in wakes behind wind turbines~\cite{sarlak2015}. Therefore, we propose considering a scale-adaptive turbulence model for the residual stress using the concept that vortex-stretching is a primary mechanism for the downscale cascade of energy. Another goal is to improve the wake model of turbines by considering that the actuation of a turbine can form a spherical canopy of stresses around the rotor.

\subsection{Related work}
LES is currently one of the most efficient turbulence modelling approaches, which has substantially impacted the remarkable growth of wind energy worldwide \cite[][]{churchfield2012,meneveau2019}. LES is of great significance in wind energy and the atmospheric science sector in studying the complex interaction of atmospheric turbulence and large wind farms  \cite[][]{Calaf2010,porte2011,porte2013,churchfield2012,xie2017}. 

In LES, a closure model for the residual stress $\tau_{ij}$ enables the accurate prediction of turbulence statistics and local energy cascade rate $\Pi(\textbf{x}, t)$. The most common parameterization for residual stresses in LES is the Smagorinsky model \cite{smagorinsky1963}. The Smagorinsky length scale is $l_{s} = C_{s}\Delta$, where $C_{s}$ is the model coefficient and $\Delta$ is the grid spacing. Existing methods for \emph{a priori}  determination of $C_s$ and its dependency on local conditions such as atmospheric stability, surface-normal distance, and mean shear are complex tasks and dependent on advanced knowledge of turbulence theory. The model coefficient need adjustments in the region of large turbulence anisotropy \cite{deardorff1980,hunt1988,horiuti1993,canuto1997,porte2004}. The dynamic Smagorinsky model is a procedure \cite{germano91} to estimate variations of the model coefficient $C_{s}$. This procedure is based on the scale-invariance assumption and requires test-filter operations and an average over the homogeneous flow direction. Averaging over a homogeneous flow direction is not reasonable in the case of complex geometries. Ref~\cite{meneveau1996} adapted the averaging along flow path lines known as Lagrangian averaging to alleviate this problem. A scale-dependent dynamic model was formulated by \cite{porte2000} to overcome the scale-invariance assumption. In the scale-dependent Lagrangian model,  the predicted logarithmic wind profiles were better than both the dynamic Smagorinsky and classical Smagorinsky models in the case of neutrally stratified atmospheric boundary layer flow over a homogeneous flat surface.

Furthermore, Refs \cite[][]{bou2005,stoll2006} evaluated the Lagrangian scale-dependent model for heterogeneous flat surfaces and, consequently, \cite{wan2007} performed the Lagrangian scale-dependent model on a two-dimensional sinusoidal hill. Predicting momentum and turbulence kinetic energy (TKE) becomes highly accurate if LES adequately resolves the vortical and straining motion in wind turbine wakes \cite[][]{sarlak2015}. As a result, choosing a subgrid-scale model may not be a determining factor in the LES of wind farms. However, in the LES of a wind farm, resolving the large eddies in the surface layer ($10\%$ of ABL) and the near-wake region of a wind turbine, {\em e.g.,} helical tip vortices, is not possible with computational resources in hand.  

The actuator disk model (ADM) and actuator line model (ALM) are the most popular methodology to parameterize turbine-induced forces. They are extensively used in large-eddy simulations of wind farms by many researchers \cite[see][]{Calaf2010,porte2011, wu2011,churchfield2012,wu2013,stevens2018}. The actuator disk model (ADM) does not require resolving the boundary layer, and thus it significantly reduces the computational cost. In contrast, the actuator line model (ALM) requires almost $40$ grid points along each line of ALM to capture the tip vortices \cite[][]{sarlak2015}. The actuator disk model only requires $5-10$ grid points across the rotor diameter. However, ADM could not produce the tip vortices at this resolution and could not fully capture the vortex shedding behind the disk, and consequently, power is over-predicted by more than $10\%$ \cite{shapiro2019}. In addition, the wind turbine structures such as blades, hub, and nacelle \cite[][]{crespo1999b}, strongly influence the flow fields in $\emph{near-wake}$,  which leads to a significantly complex, three-dimensional turbulent flow in this region.

On the other hand, the flow field in the $\emph{far-wake}$ is less likely to have a significant influence from wind turbine structures. Thus, the actuator disk theory predicts the \emph{far-wake} mean flow field with the global parameters such as momentum, power coefficient, and inflow condition. Researchers proposed several modifications to improve the actuator disk model; for example,  \cite[][]{wu2011} used a modified ADM, which applies thrust and torque to the incoming flow. The results of this modified ADM were in good agreement with the experimental observations. However, this approach requires {\em a priori} knowledge of lift and drag coefficients. For ADM and ALM, Ref \cite{stevens2018} also investigated the influences of the wind turbine structures, {\em e.g.,} nacelle and tower. Numericals tests of ADM and ALM incorporating structural effects in the LES indicate an excellent correlation with the experimental data.  Nevertheless, it increases the complexity of the numerical simulation.
\subsection{Outline}
The organization of this article is as follows. First, Sec \ref{sec:NM} provides a brief outline of the fundamentals of the parameterization of turbulence and turbines and lays a general-purpose formulation of the algorithms, making them applicable to any existing research LES code.  Next, in Sec \ref{sec:RA}, we summarize the findings of our numerical simulations: an isolated wind turbine and an array of $30$ wind turbines. In particular, we employ wind tunnel measurements to understand how to extend the actuator disk theory for accurate modeling of wind turbine wakes. Finally, Sec~\ref{sec:cfd} summarizes and discusses key findings and highlights potential future extensions, some of which are currently underway.

\section{Gaussian actuator disk model for LES of wind farms}
\label{sec:NM}
\subsection{Large eddy simulation framework}
The filtered version of continuity and momentum equations are:
\begin{equation}
	\frac{\partial \bar u_{i}}{\partial x_{i}} = 0,
\end{equation}

\begin{equation}
	\frac{\partial \bar{u_{i}}}{\partial t} + \bar u_{j} \frac{\partial u_{i}}{\partial{ x_{j}}} = -\frac{\partial \bar{p}}{\partial x_{i}} - \frac{\partial \tau_{ij}}{\partial x_{j}} + f_{i} \chi(\textbf{x},t),
	\label{eq:nse}
\end{equation}
where $\bar{u}_{i}$ ($i = 1, 2, 3$) is the filtered velocity component in the streamwise ($x_{1}$), spanwise ($x_{2}$) and surface-normal direction ($x_{3}$), respectively. The function $\chi(\textbf{x}, t)$ is either 0 or 1 depending on where the body force is applied. The residual (sub-filter) stress tensor $\bar\tau_{ij} = \overline{u_{i}u}_{j}- \bar u_{i}\bar u_{j}$, accounts for the effects of small-scale motions smaller than the filter width $\Delta_{\hbox{\tiny{les}}}$. The following section discusses parameterization techniques for the subgrid-scale turbulence and wind turbine.

\subsection{Turbulence modelling} 
\subsubsection{Smagorinsky model}
Among various models for subgrid-scale turbulence, the Smagorinsky model is used widely~\cite[][]{smagorinsky1963}, where the stress tensor $\tau_{ij}$ appeared in Eq (\ref{eq:nse}) is represented as
\begin{equation}
	\tau_{ij}-\frac{1}{3}\tau_{kk}\delta_{ij}=-2\nu_{\tau}\mathcal{\bm S}_{ij}.
	\label{eq:stresssub}
\end{equation}
Here, $\nu_{\tau} (\bm{x}, t) = C_{s}\Delta_{\tiny{les}}^{2}(2\mathcal{\bm S}_{ij}\mathcal{\bm S}_{ij})^{1/2}$ is the eddy viscosity, $\delta_{ij}$ represents kronecker delta and $\mathcal{S}_{ij} = 1/2 \left(\mathcal{G}_{ij} + \mathcal{G}_{ji}\right)$ is the symmetric part of the velocity gradient tensor $\mathcal{G}_{ij}=\partial u_i/\partial x_j$ and the model coefficient $C_{s}=(1/\pi)(3K_c/2)^{-3/4}$ is known as the Smagorinsky constant. For a Kolmogorov constant of $K_c \approx 1.4$, one gets $C_s \approx 0.18$. At high Reynolds number, the production of vortices becomes highly intermittent in space and time \cite{germano91}. It is thus necessary to consider the Leonard (or resolved stress) $\tau^{L}_{ij} = \widetilde{\bar{u}_{i}\bar{u}_{j}}-\tilde{\bar{u_{i}}}\tilde{\bar{u_{j}}}$ to introduce spatio-temporal variability of the Smagorinsky coefficient $C_{s}(\bm x, t)$. This approach is known as the dynamic Smagorinsky model, in which $C_{s}(\bm x, t)$ becomes negative, thereby leading to numerical instability. However, several sophisticated models exists as a workaround, making the dynamic Smagorinsky model a reliable scheme. Note that the energy flux of the dynamic Smagorinsky model takes the following form:
\begin{equation}
	\bm \Pi = C_{s}^{2}(\bm x, t)\Delta_{\hbox{\tiny les}}^{2} 2^{3/2}(\mathcal{\bm S}_{ij} \mathcal{\bm S}_{ij})^{3/2},
	\label{eq:fluxSmagorinsky}
\end{equation}
Clearly, the dynamic Smagorinsky model does not directly account for the production of vorticity in wind turbine wakes and the associated intermittency of turbulence. It is important to note that the tip-speed ratio, $\lambda = |\bm\omega|D/(4u_d)$, of a turbine of rotor diameter ($D$), relates the vorticity ($\bm\omega$) with the local fluid velocity $u_d$. Thus, the rotation of turbines influences the momentum conservation law {\em via} the vortex stretching term $\mathcal{\bm S}_{ij}\omega_{j}$ that appears in the vorticity form of the Navier-Stokes equation. Consequently, wind turbines contribute to the dynamical evolution of the enstrophy ($0.5|\bm\omega|^2$) {\em via} vorticity stretching $\omega_{i}\mathcal{\bm S}_{ij}\omega_{j}$. Readers may find a technical details of vortex stretching theory in Chapter~5 of Ref~\cite{davidson2015turbulence}. Below, we present our development of a subgrid model that accounts for the effects of vortex stretching.
\subsubsection{Vortex-stretching based scale-adaptive subgrid model}
Consider the truncated Taylor-series expansion
\begin{equation}
	\bar u_{i}(\bm{x}, t) = \tilde{\bar u}_{i}(\bm{r}, t) + (\bm x-\bm r)\mathcal{G}_{ij},
	\label{eq:taylor}
\end{equation}
where $\bm x = \bm r + \alpha\Delta_{\hbox{\tiny les}}$. Assuming the solution of Navier-Stokes equations remains smooth and continuous at high Reynolds number, one can express the Leonard stress approximately as
\begin{equation}
	\tau^{L}_{ij} = c_{k}\Delta^{2}_{\hbox{\tiny{les}}}\mathcal{G}_{ik}\mathcal{G}_{jk}.
	\label{eq:Leonard}
\end{equation}
In the literature, Eq~(\ref{eq:Leonard}) is known as the gradient model, a complete discussion of which is outside the scope of the current research. Using Eq (\ref{eq:Leonard}), we get the following form of the energy flux,
\begin{equation}
	\bm \Pi =C_{k}\Delta^{2}\left[-\mathcal{S}_{ik}\mathcal{S}_{kj}\mathcal{S}_{ji} +\frac{1}{4}\omega_{i}\omega_{j}\mathcal{S}_{ij}\right],
	\label{eq:fluxvortst}
\end{equation}
where $\omega_{i} = \epsilon_{ijk}\mathcal{G}_{kj}$ is the vorticity vector and $\mathcal S_{ik}\mathcal S_{kj}\mathcal{S}_{ji}$ is the skewness of the filtered strain tensor. Comparing Eq (\ref{eq:fluxSmagorinsky}) and Eq (\ref{eq:fluxvortst}), one finds that the use of Eq (\ref{eq:Leonard}) as a subgrid model would make the energy flux depend directly on strain skewness and vorticity stretching.
\par After some algebraic manipulation, we can write
\begin{equation}
	-\frac{1}{2}\tau^{L^{dev}}_{ij}\tau^{L^{dev}}_{ji} = -\frac{1}{4}\left[\mathcal{S}_{ij}\omega_{j}\mathcal{S}_{ik}\omega_{k} + \frac{1}{3}(\mathcal{G}_{ij}\mathcal{G}_{ji})^2 \right].
	\label{eq:LeonarDev}
\end{equation}
Following Deardorff's modification of Smagorinsky model, the subgrid scale stress in Eq (\ref{eq:stresssub}) can be given by
\begin{equation}
	\label{eq:tau}
	\tau_{ij} - \frac13\tau_{kk}\delta_{ij} = c_k\Delta_{\hbox{\tiny les}}\sqrt{k_{\hbox{\tiny sgs}}}\mathcal S_{ij}.
\end{equation}
Based on dimensional consideration, an algebraic model for the subgrid-scale turbulence kinetic energy may take the following form:
\begin{equation}
	\label{eq:ksgs}
	k_{\hbox{\tiny sgs}} = \frac{\Delta^2_{\hbox{\tiny les}}\left(\frac12\mathcal S_{ij}\omega_j\mathcal S_{ik}\omega_k + \frac16(\mathcal G_{ij}\mathcal G_{ji})^2\right)^3}{\left[(\mathcal S_{ij}\mathcal S_{ij})^{5/2} +(\frac12\mathcal S_{ij}\omega_j\mathcal S_{ik}\omega_k + \frac16(\mathcal G_{ij}\mathcal G_{ji})^2)^{5/4}\right]^2}.
\end{equation}
Note that the subgrid model Eq (\ref{eq:tau}) implicitly adapts the energy flux as the characteristic length scale of turbulence dissipation varies locally. Recent work of \cite{Hossen2021, bhuiyan2020, Alam2018} evaluated the scale-adaptive LES model, Eq (\ref{eq:tau}), in the case of isotropic turbulence, and atmospheric boundary layer flow over urban and hilly terrain. In this research, we consider an improved actuator
disk model and show that our scale-adaptive LES method adequately captures the near turbine dynamics.

To approximate the parameter $c_k$, we consider fixed values of the Smagorinsky constant $$C_s(\bm x,t) = \frac1\pi\left(\frac{3K_c}{2}\right)^{-3/4}$$ and the Kolmogorov constant $K_c=1.4$ in Eq~(\ref{eq:fluxSmagorinsky}). Then, we obtain $c_k$ by equating the energy flux given by Eq~(\ref{eq:fluxSmagorinsky}) to the energy flux obtained by employing Eq~(\ref{eq:tau}). This is equivalent to mimicking the effects of dynamical variations of $C_s(\bm x,t)$ {\em via} the stretching of vortices. Based on LES of decaying homogeneous isotropic turbulence for Reynolds numbers up to $10^8$ (using the localized dynamic Smagorinsky model~\cite{Hossen2021}), we found that $c_k = 0.325\pm0.035$. Thus, a fixed value of $c_k=0.325$ is considered in this article.

\subsection{Wind turbine parameterization}
In this section, we extend classical ADM by considering that the actuation of a turbine forms a spherical canopy of stresses around the rotor. The actuator disk theory considers the momentum principle in a stream tube around a turbine, in which the thrust force is
\begin{equation}
	f = -\frac{1}{2}\rho C{'}_{t} u_{d}^{2}A_{c}.
	\label{eq:actuatorforce}
\end{equation}
This thrust is applied as a momentum sink due to fluid elements passing over the rotor disk. The streamwise velocity $u_{d}$ is averaged over the actuator disk and related to the upstream undisturbed velocity ($U_{\infty}$) such that $u_{d} = (1-a)U_{\infty}$, where $a$ is the axial induction factor, $A_{c}$ is the frontal area, and $C{'}_{t} = C_{t}/(1-a)^{2}$ is a modified form of the thrust coefficient $C_t$. This model leads to a flattened near-wake profile because it does not account for the wind speed variation across the rotor. We have considered the following two developments. For a pedagogical reason, these two formulations will be referred to as `model A’ and `model B’, respectively. 

According to the blade element momentum theory, the power coefficient $C_p=4a(1-a)^2$ accounts for the wind power efficiency. However, the dynamic load is related to the turbulent intensity in wakes~\cite{chamorro2009}. Thus, in model A, we consider the following form of the thrust force
\begin{equation}
	f_{i}(\bm x, t) = -\frac{1}{2}\rho C{'}_{t} A_{c}u_{d}\bar{u}_{i}(\bm x, t),
	\label{eq:diskavg}
\end{equation}
in which the instantaneously resolved velocities $\bar{u}_{i}(\bm x, t)$ in all three directions are considered in Eq (\ref{eq:diskavg}). This force is projected onto the numerical domain using a three-dimensional Gaussian kernel 
$$
G(r) = \frac{1}{\sqrt{2\pi}\Delta_G}  \exp \left(-\frac{1}{2}\frac{r^2} {\Delta^{2}_{G}}\right),
$$
where $r$ represents distance from the centre of the actuator disk to the point where the influence of turbine is assumed, $\Delta_{G}$ is the width of Gaussian kernel. Then the total thrust force ($F_{T}$) takes the following form
\begin{equation}
	F_{T} = \frac{1}{\Delta v} \left[f_{i}(\bm x, t) \circledast G(r)\right],
	\label{GaussianEq}
\end{equation}
where the symbol `$\circledast$' is the convolution between the two functions, and $\Delta v$ is the volume of a grid cell.

\begin{figure}[h!]
	\centering
	\includegraphics[scale=0.5]{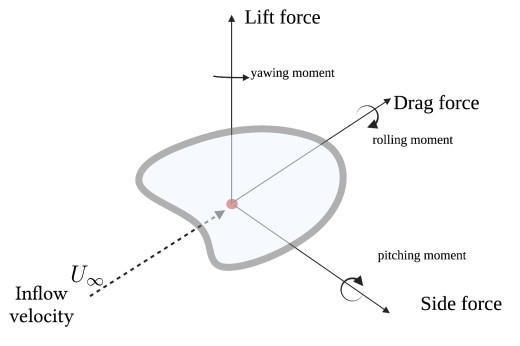}
	\caption{Schematic diagram illustrates the forces and moments acting on an arbitrary object when immersed in a fluid flow.}
	\label{fig:bodyforce}
\end{figure}

In model B, we consider the fluid dynamics point of view in which a solid body of an arbitrary shape is immersed in a fluid, as shown in Fig \ref{fig:bodyforce}. An immersed body will experience forces and moments from all three principal axes. Let us assume that the body is immersed in a fluid flow, and the incoming flow stream is parallel to the main chord line of the body and has symmetry about the lift-drag axis such as cylinder, sphere, wings, etc. The body will then only experience a drag force. This approach considers the momentum principle in a control volume around a wind turbine. Considering the loss of momentum and pressure due to frictional effects, we arrive at the following form of total drag experienced by the turbine:
\begin{equation}
	f_{i}(\bm x,t) = -C_{\nu}\bar{u}_{i}(\bm x,t) - C{'}_{t}\lvert \bar{\bm u} \rvert \bar{u}_{i}(\bm x,t),
	\label{eq:actuationEQ}   
\end{equation}
where $|\bar{\bm u}| = \bar u_i\bar u_i$ is a function of $(\bm x,t)$.
Here, $C_{\nu}$ is the viscous drag coefficient, and $C{'}_{t}$  is the coefficient of pressure drag. We have applied the thrust force (~given by Eq \ref{eq:actuationEQ}~) over a spherical canopy with a center that coincides with the center of a rotor similarly as in model-A but without using a Gaussian kernel. 

\section{RESULT ANALYSIS}
\label{sec:RA}
The scale-adaptive LES framework discussed in this article has been implemented in an updated version of the (collocated grid) Navier-Stokes solver that was detailed previously by \cite{Alam2018}. Previous utilization of the code investigated the interaction of geophysical turbulent flow over an array of wind turbines using classical ADM \cite[][]{alam2019}, as well as turbulent flow over the Askervein hill, UK \cite[][]{bhuiyan2020}. Here, we investigate the new development of the Gaussian actuator disk model of the wind turbine and compare the numerical results against the experimental data.

\begin{figure}[htb!]
	\centering
	\includegraphics[scale=0.32]{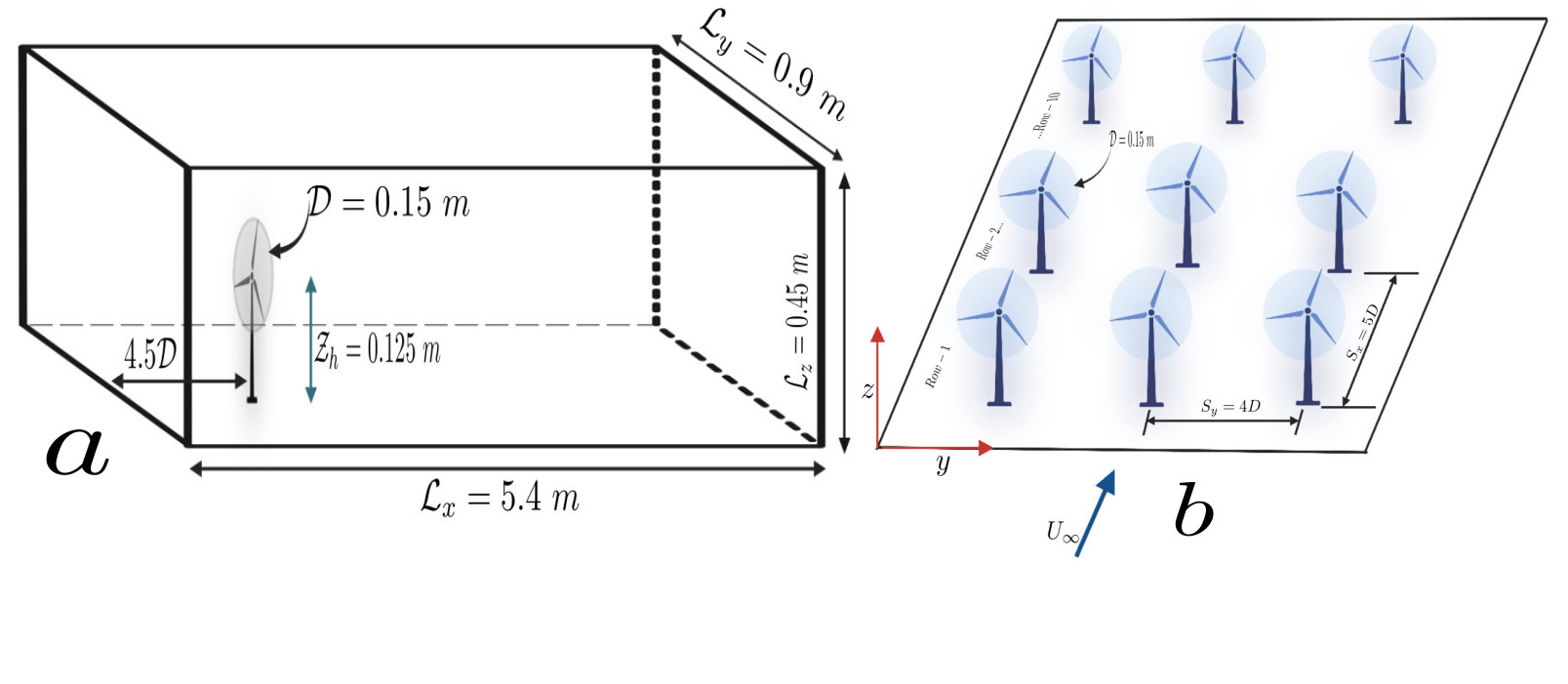}
	\caption{\textbf a) The computational domain $\mathcal{L}_{x} \times \mathcal{L}_{y} \times \mathcal{L}_{z} = [5.4 \times 0.9 \times 0.45$]~m$^{3}$, which represents the simulated part of the wind tunnel test section. \textbf b) The layout of an array of $10\times 3$ wind turbines in the computational domain $\mathcal{L}_{x} \times \mathcal{L}_{y} \times \mathcal{L}_{z} = [14.4 \times 1.8 \times 0.675$]~m$^{3}$. Only first two and the last row of the array $10 \times 3$ are shown for clarity. The streamwise and spanwise spacing between wind turbines are $\mathcal{S}_{x} = 5D$ and $\mathcal{S}_{y} = 4D$, respectively.}
	\label{fig:setup}
\end{figure}
\begin{figure}[h!]
	\centering
	\includegraphics[scale=0.22]{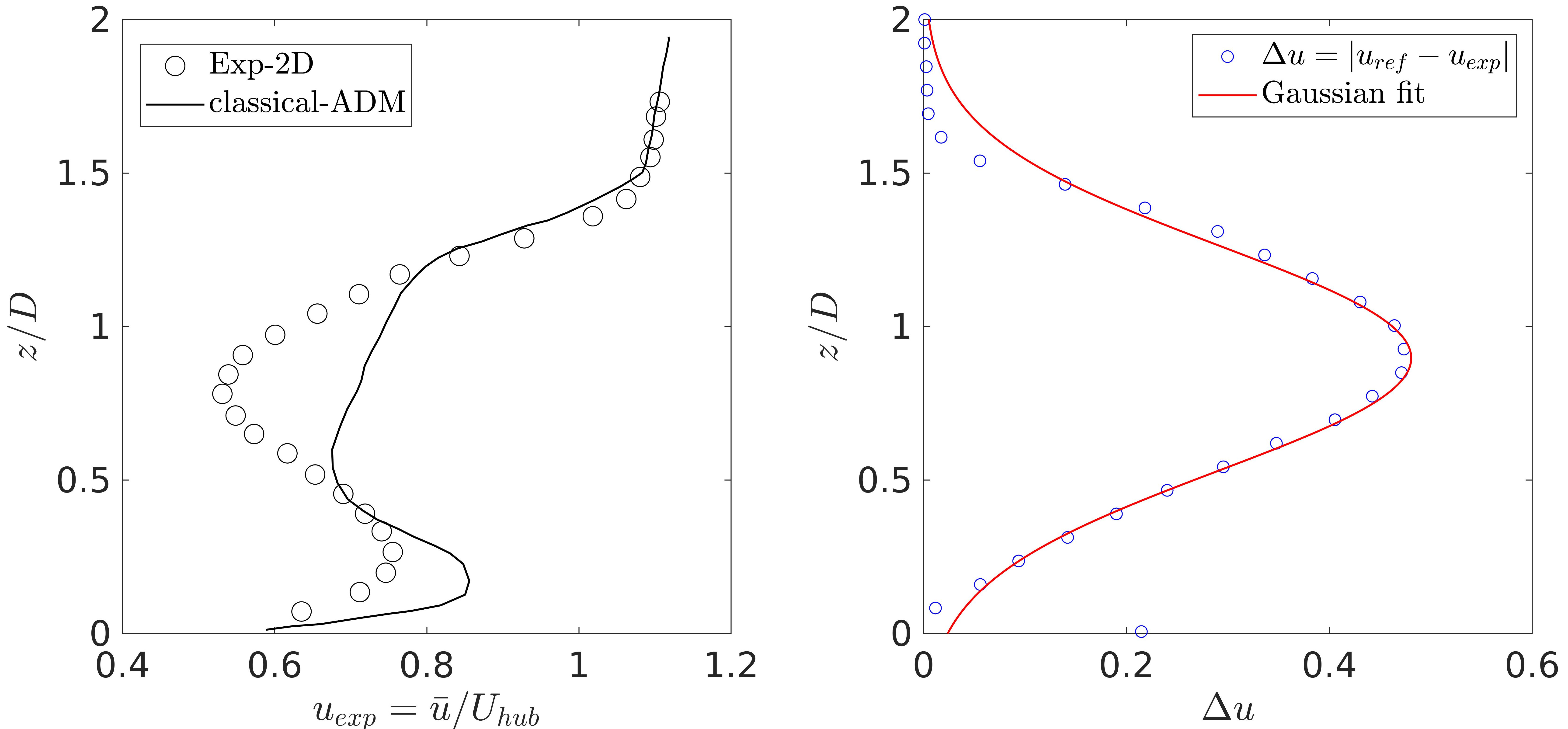}
	\caption{\textbf a) A comparison of the wind tunnel measurement at $x/D=2$ with the corresponding LES result of~\cite{stevens2017} using classical-ADM. \textbf b) A comparison of velocity deficit in the experimental data \cite{chamorro2010} with a Gaussian curve.}
	\label{fig:gaussianST}
\end{figure}

Fig \ref{fig:setup}a-b represents the test sections of the wind-tunnel experiment \cite{chamorro2010,chamorro2011}, which have been simulated in the present work. The rotor diameter and the hub height of the turbine are $0.15$~m and $0.125$~m, respectively. The data of the first experiment consists of the velocity profiles (and Reynolds stresses) on the vertical mid-plane at 8 locations: $x/D = -1$, $x/D = 2$, $x/D = 3$, $x/D = 5$, $x/D = 7$, $x/D = 10$, $x/D = 14$, $x/D = 20$, where $x/D=0$ is the center of the turbine which is located at $4.5D$ from the inlet plane. Fig \ref{fig:gaussianST}a compares the vertical profile of the streamwise velocity between the experimental data at $x/D = 2$ and the corresponding LES result using classical ADM \cite{stevens2017}. As discussed in the introduction, the wind profile from the classical ADM has been flattened. More specifically, sharp deviation is seen in the boundary layer below the rotor bottom. Fig \ref{fig:gaussianST}b shows the velocity deficit ($\Delta U = \lvert u_{ref}-u_{exp}\rvert$) of the experimental data at $x/D = 2$, where $u_{ref}$ is the streamwise velocity at the inlet of the wind-tunnel test section. Using the curve-fitting tool of \texttt{MATLAB}, we observe that the experimental velocity deficit shows a perfect agreement with a Gaussian profile (Fig \ref{fig:gaussianST}b). The experimental data suggest the validity of the assumptions of the Gaussian profile of the thrust across the actuator disk in model A. This observation is particularly relevant to generalizing the velocity deficit in classical non-rotating ADM and developing a model that accounts for the rotational effects of the disk~\cite{chamorro2009}.

The second set of experimental data is velocity profiles (and Reynolds stresses) on the vertical mid-plane of the wind tunnel test section for a $10\times 3$ array of wind turbines at 4 locations: $x/D=1, x/D=2, x/D=3, x/D=4$ behind each row of wind turbines. A Gaussian pattern of the near turbine wake was also observed for this set of experimental data.

\subsection{Gaussian actuator disk model}
A parametric study of the Gaussian actuator disk model was considered for 11 test cases (see Table \ref{tab:allcases}). In these tests, the aspect ratio of the horizontal grid spacing to the vertical grid spacing was varied. 
\subsubsection{Mesh sensitivity analysis with an isolated turbine}
\label{sec:ABLST}
\begin{figure}[b!]
	\centering
	\includegraphics[scale=0.32]{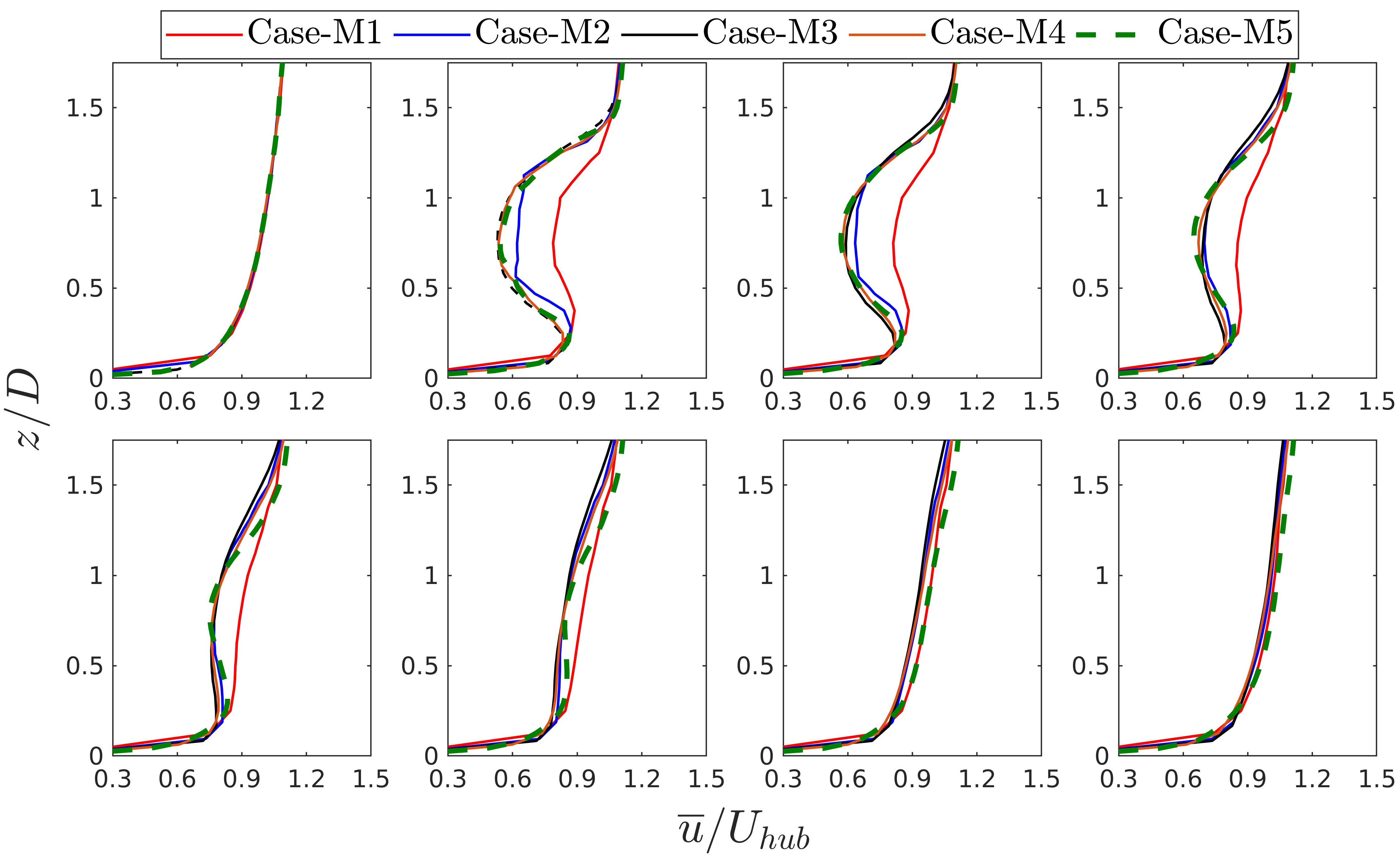}
	\caption{A Comparison of vertical profiles of mean streamwise velocity for five different mesh resolutions in case of model A (see, Table \ref{tab:allcases}).}
	\label{fig:M1}
\end{figure}

\begin{figure}[htb!]
	\centering
	\includegraphics[scale=0.3]{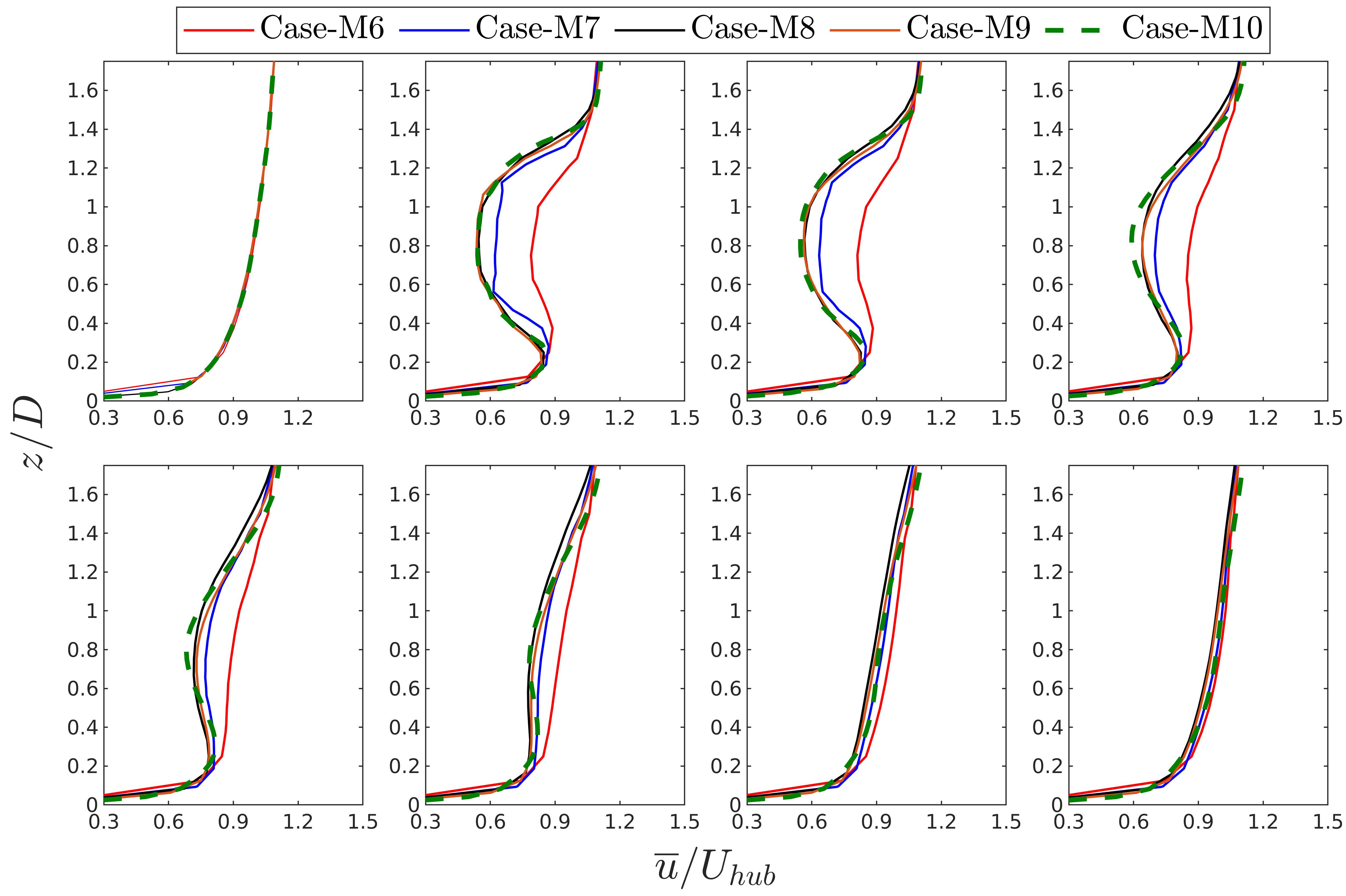}
	\caption{\label{M2}A comparison of vertical profiles of mean streamwise velocity for five different mesh resolution in case of model B (see, Table (\ref{tab:allcases})).}
	\label{fig:M2}
\end{figure}

\begin{table}[b!]
	\centering
	\caption{\label{tab:allcases} Numerical setup of all the cases used in the large eddy simulations with different grid resolution for model A, model B and classical ADM. Case M1 to M11 are for single wind turbine while case M12, M13, and M14 are used for wind farm simulation. Here, $\Delta = (\Delta x \Delta y \Delta z)^{1/3}$ and $\Delta_{G}$ is the width of Gaussian kernel. $N_{x_{2}}$ and $N_{x_{3}}$ represents the number of grid points in the spanwise and surface-normal direction of the rotor.}
	\begin{tabular}{llccccc}
		\hline
		Cases  &Model &Resolution &$N_{x_{2}}$
		&$N_{x_{3}}$ &$\Delta_{G}/D$ &$\Delta/D$  \\
		\hline
		case M1 & model A & $48\times 9 \times 16$ & $1$ & $5$
		& $3.23$ & $0.45$  \\
		case M2 & model A & $64\times 16 \times 24$ & $3$ & $8$
		& $3.23$ & $0.29$\\
		case M3 & model A & $128\times 21 \times 36$ & $3$ & $12$
		& $3.23$&$0.18$ \\
		case M4 & model A & $180\times 30 \times 48$ & $5$ & $16$
		& $3.00$&$0.13$ \\
		case M5 & model A & $324 \times 56 \times 72$ & $9$ & $24$
		& $3.00$&$0.079$ \\
		case M6 & model B & $48\times 9 \times 16$ & $1$ & $5$
		& $-$ & $0.45$  \\
		case M7 & model B & $64\times 16 \times 24$ & $3$ & $8$
		& $-$ & $0.29$\\
		case M8 & model B & $128\times 21 \times 36$ & $3$ & $12$
		& $-$&$0.18$ \\
		case M9 & model B & $180 \times 30 \times 48$ & $5$ & $16$
		& $-$&$0.13$ \\
		case M10 & model B & $324 \times 56 \times 72$ & $9$ & $24$
		& $-$&$0.079$ \\
		case M11 & classical ADM & $180 \times 30 \times 48$ & $5$ & $16$
		& $1.46$&$0.13$ \\
		case M12 & classical ADM & $1024 \times 128 \times 144$ & $10$  &$32$
		& $1.45$ & $0.065$ \\
		case M13 & model A & $476 \times 60 \times 72$ & 5  & 16
		&2.40& $0.13$ \\
		case M14 & model B & $476 \times 60 \times 72$ & 5  & 16
		&--& $0.13$ \\
	\end{tabular}
\end{table}
%
Table \ref{tab:allcases}  reports the number of grid points $N_{x_{2}}$ and $N_{x_{3}}$ across the actuator disk in spanwise and vertical directions, respectively. This study considers a decreasing sequence of average grid spacing $\Delta/D$ in the range $[0.45, 0.079]$ for both model A and model B (see, Table \ref{tab:allcases}). Numerical data were sampled at eight streamwise locations following the aforementioned wind-tunnel experiment. Vertical profiles of the mean streamwise velocity  are shown in Fig \ref{fig:M1} and Fig \ref{fig:M2} for model A and model B, respectively. Flattened profiles are seen only for the two coarsest resolution cases: (M1, M2) for model A and (M6, M7) for model B. This behaviour is quite similar to the predictions made with the classical ADM ({\em e.g.}, see Refs \cite{stevens2017} and \cite{stevens2018}). The velocity profiles for cases (M3-M5 and M8-M10) with the most refined grid exhibit an excellent mutual agreement. More specifically, the LES results for model A (cases, M3-M5) show negligible sensitivity to the mesh resolution, thereby indicating that momentum fluxes converge in both the near-wake ($x/D < 5$) and the far-wake ($x/D>5$) region. A similar trend in the wake predicted by model B suggests that frictional effects of the blades and nacelle can be averaged. However, two-way feedback between the turbine wake and the atmosphere is a primary factor improving the actuator disk model.
Interestingly, there is only one grid point across the rotor in case M1 and case M6. Nevertheless, the trend in the velocity profile closely resembles the experimental velocity profile in the near-wake region. Also, it indicates that a dynamic approach of adjusting the thrust coefficient could correct errors at the coarsest resolution profiles. Indeed, Ref~\cite{roy2011simulating} had only one grid point across the rotor in which the horizontal grid spacing was $1$~km. In cases M2 and M7, there are only three grid points across the rotor in the horizontal direction. The observations suggest that a reinforcement machine learning algorithm maybe developed for adjusting the thrust coefficient dynamically whenever a sufficiently fine grid spacing is not affordable. Thus, our Gaussian ADM coupled with the scale-adaptive turbulence model has the potential to provide relatively more accurate quantitative understanding of the effect of wakes in annual energy production of actual large wind farms. Below, we compare the velocity profiles of the Gaussian ADM (cases M4 and M9), classical ADM (case M11), and wind-tunnel measurement.

\subsubsection{Validation of the Gaussian ADM against wind tunnel measurements}
\begin{figure}[htb!]
	\centering
	\includegraphics[scale=0.35]{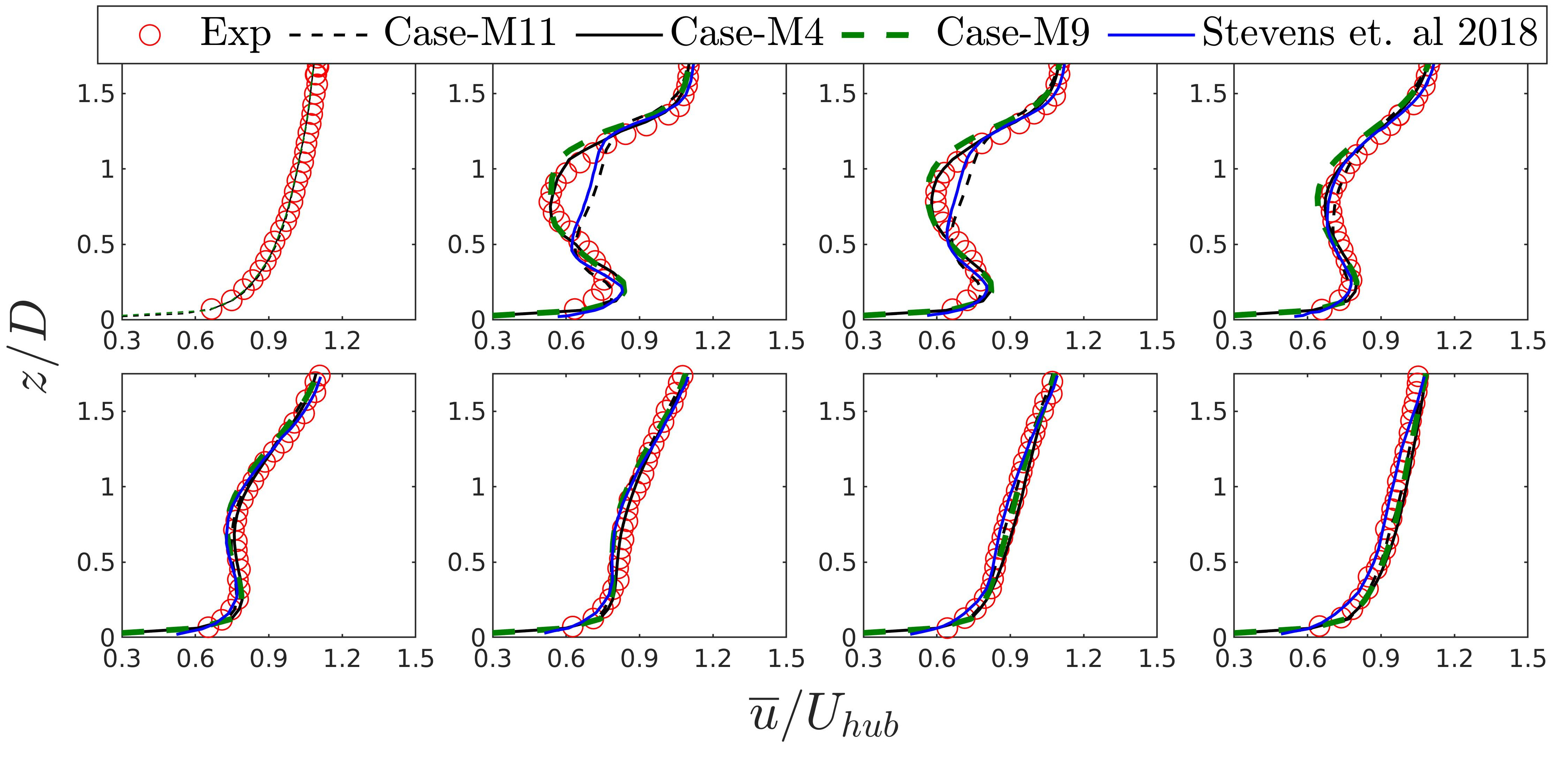}
	\caption{\label{ch1-fig3} A comparison of vertical profiles of mean streamwise velocity among wind tunnel measurements \cite[][]{chamorro2010}, corresponding LES data \cite[][]{stevens2018}, and our present LES results for classical ADM, model A and model B.}
	\label{stmean1}
\end{figure}

In Fig \ref{stmean1}, we compare the mean streamwise velocity profiles of present LES, the wind tunnel data \cite[][]{chamorro2010}, and the LES data provided by Ref~\cite{stevens2018}. When we look at the mean velocity profiles obtained from case M4 and case M9, we see that the result of the Gaussian ADM has a good agreement with the experimental data in the \emph{far-wake} ($x/D > 5$) region, which agrees with the outcome of classical ADM that appeared in the literature \cite[][]{jimenez2007,wu2011,porte2011,wu2013,yang2013,yang2014}. However, the Gaussian ADM also accurately captures the velocity distribution in the \emph{near-wake} region ($x/D<5$). Fig \ref{stmean1} reveals that the actuation of horizontal axis wind turbine acts as a spherical canopy of frictional and pressure drag. Experiments of wind turbines can provide limited measurement results such as velocity and pressure. Our scale-adaptive LES coupled with model A or model B indicates a balance of the efficiency and accuracy in predicting the annual energy production of modern large wind farms.

\begin{figure}[bp]
	\centering
	\includegraphics[scale=0.32]{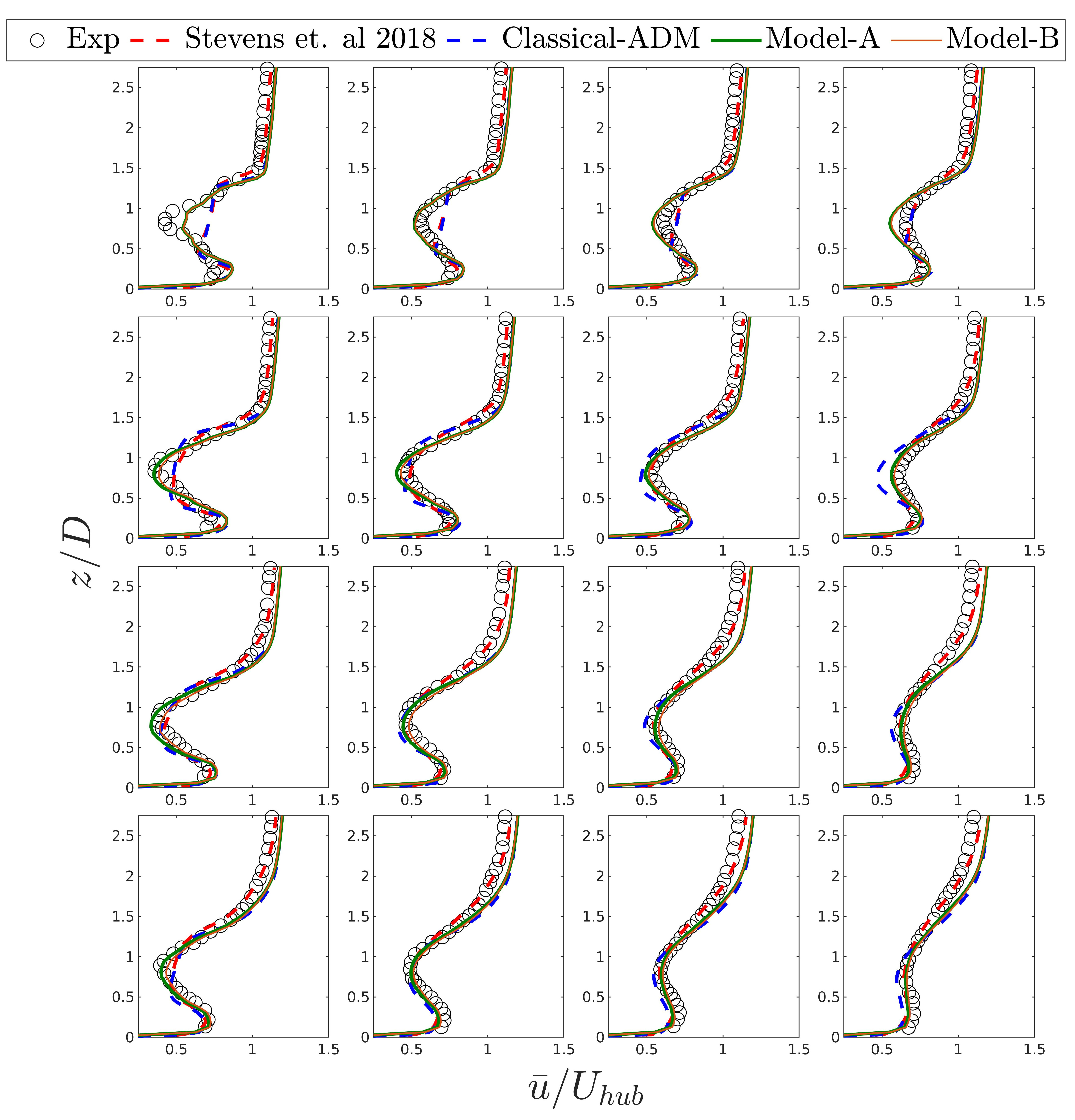}
	\caption{A comparison of vertical profiles of mean streamwise velocity between wind-tunnel measurements \cite[][]{chamorro2010}, corresponding LES data \cite[][]{stevens2018} and our present LES for classical~ADM, model~A, and model~B in case of wind farm.}
	\label{fig:wf1} 
\end{figure}
\begin{figure}[h]
	\centering
	\includegraphics[scale=0.32]{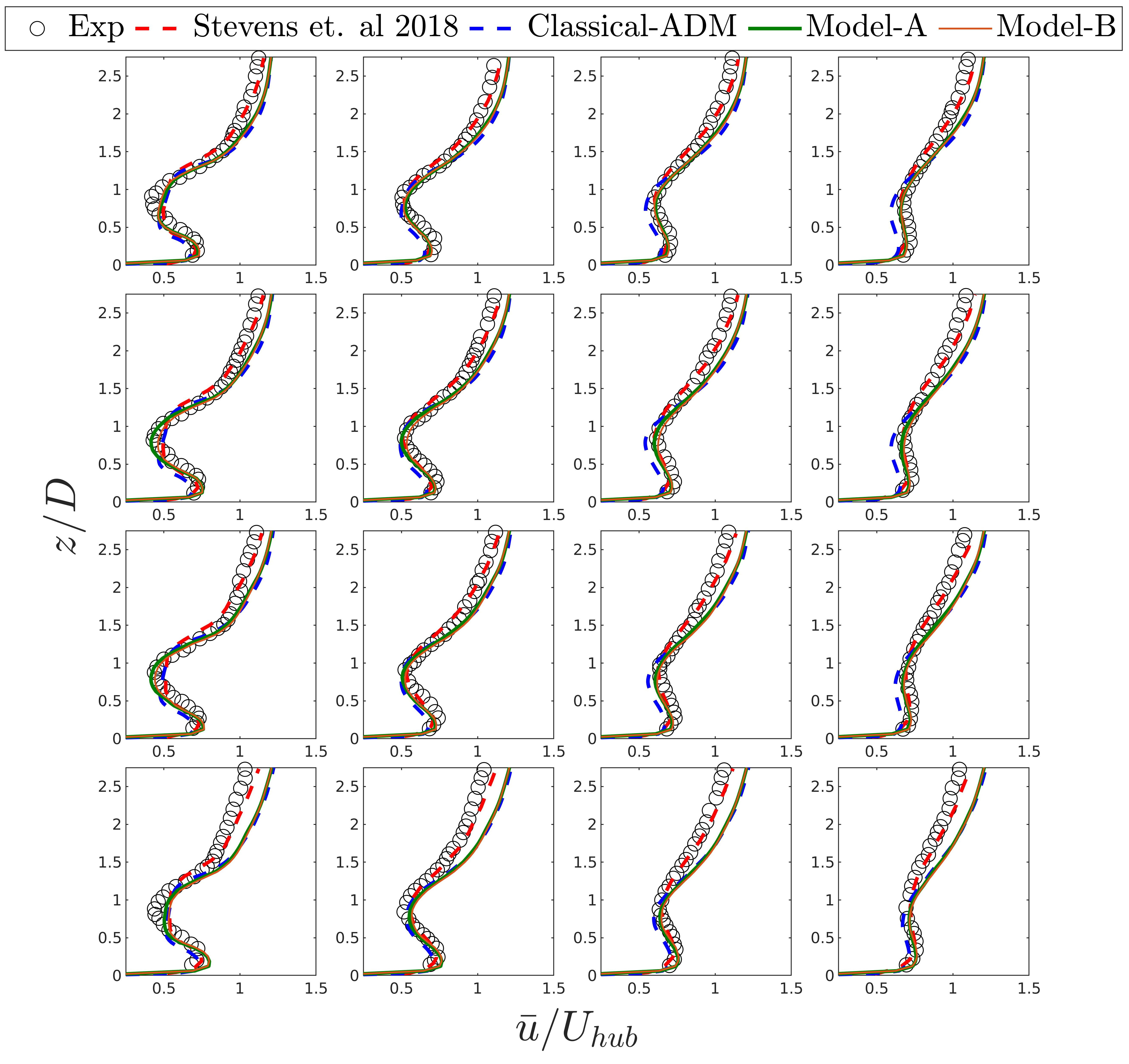}
	\caption{A comparison of vertical profiles of mean streamwise velocity between wind-tunnel measurements \cite[][]{chamorro2010}, corresponding LES data \cite[][]{stevens2018} and our present LES for classical~ADM, model~A, and model~B in case of wind farm.}
	\label{fig:wf2}
\end{figure}

The interaction of wakes  down-wind of a turbine shows a velocity deficit reducing $10\%$–$20$\% energy production. Here, we discuss the interaction of neutrally stratified atmospheric boundary layer flow over an array of $30$ wind turbines by comparing the results of scale-adaptive LES with wind tunnel measurements. It is worth mentioning that wind tunnel measurements are subject to the validity of the Reynolds number similarity theory. However, a wind tunnel validation for physical and dynamical considerations in the scale-adaptive LES of Gaussian ADM helps to understand accurate numerical design of actual wind farms. Figs \ref{fig:wf1} \& \ref{fig:wf2} show the mean streamwise velocity distribution in the vertical mid-plane of the wind farm at locations, $x/D=1, 2, 3, 4$, behind the wind turbines. Here, we have compared the velocity profiles of scale-adaptive LES with wind tunnel measurements. We have also employed the classical ADM within our scale-adaptive LES framework and compared the corresponding results with scale-dependent Lagrangian LES coupled with the classical ADM~\cite{stevens2018}. We find that the velocity profiles obtained with model A and model B have a good agreement with experimental observations \cite{chamorro2011} in the \emph{near-wake} as well as in the \emph{far-wake} region.

On the other hand, results with classical ADM are as good as previous LES data \cite{stevens2018}. We also noticed that the mean profiles obtained with classical ADM significantly improved in the \emph{near-wake} region compared to a standalone wind turbine. This is because  assumptions made in ADM closely agree with the wake-layer profile created at the hub height in the large wind farms. This finding is well aligned with the literature \cite{stevens2018}. 

\begin{figure}[t]
	\centering
	\begin{tabular}{ccc}
		\includegraphics[scale=0.17]{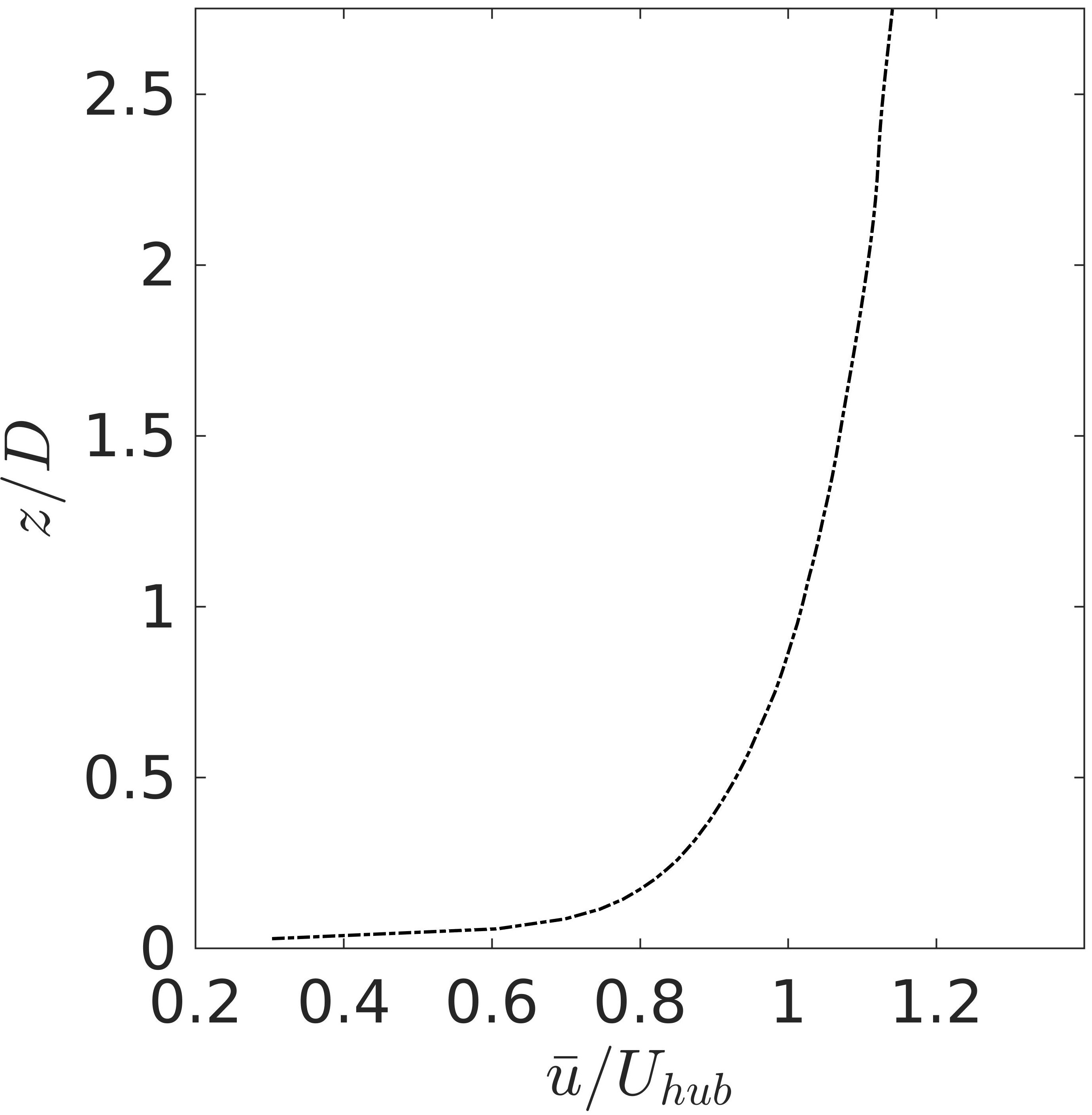}&
		\includegraphics[scale=0.18]{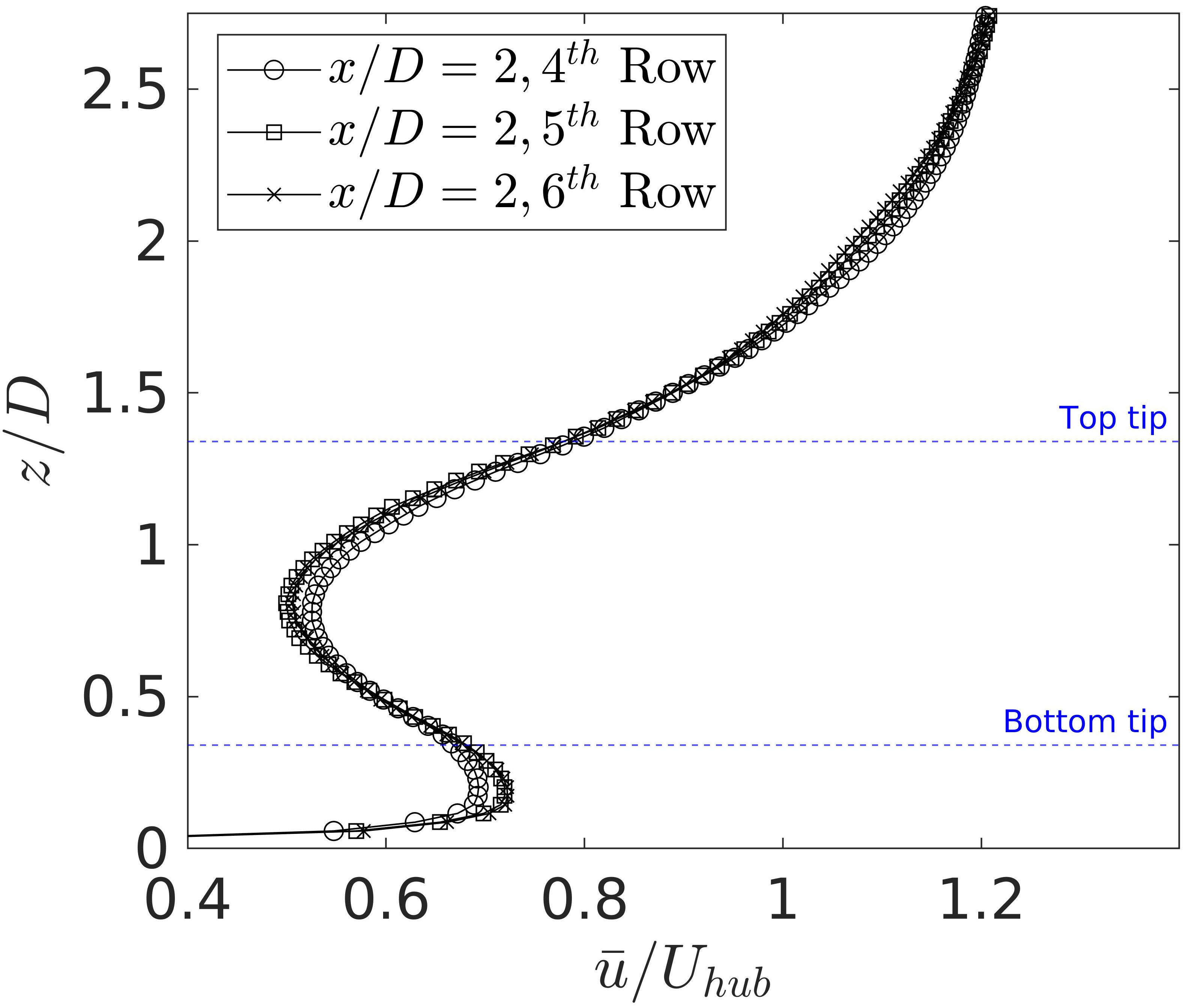}&
		\includegraphics[scale=0.185]{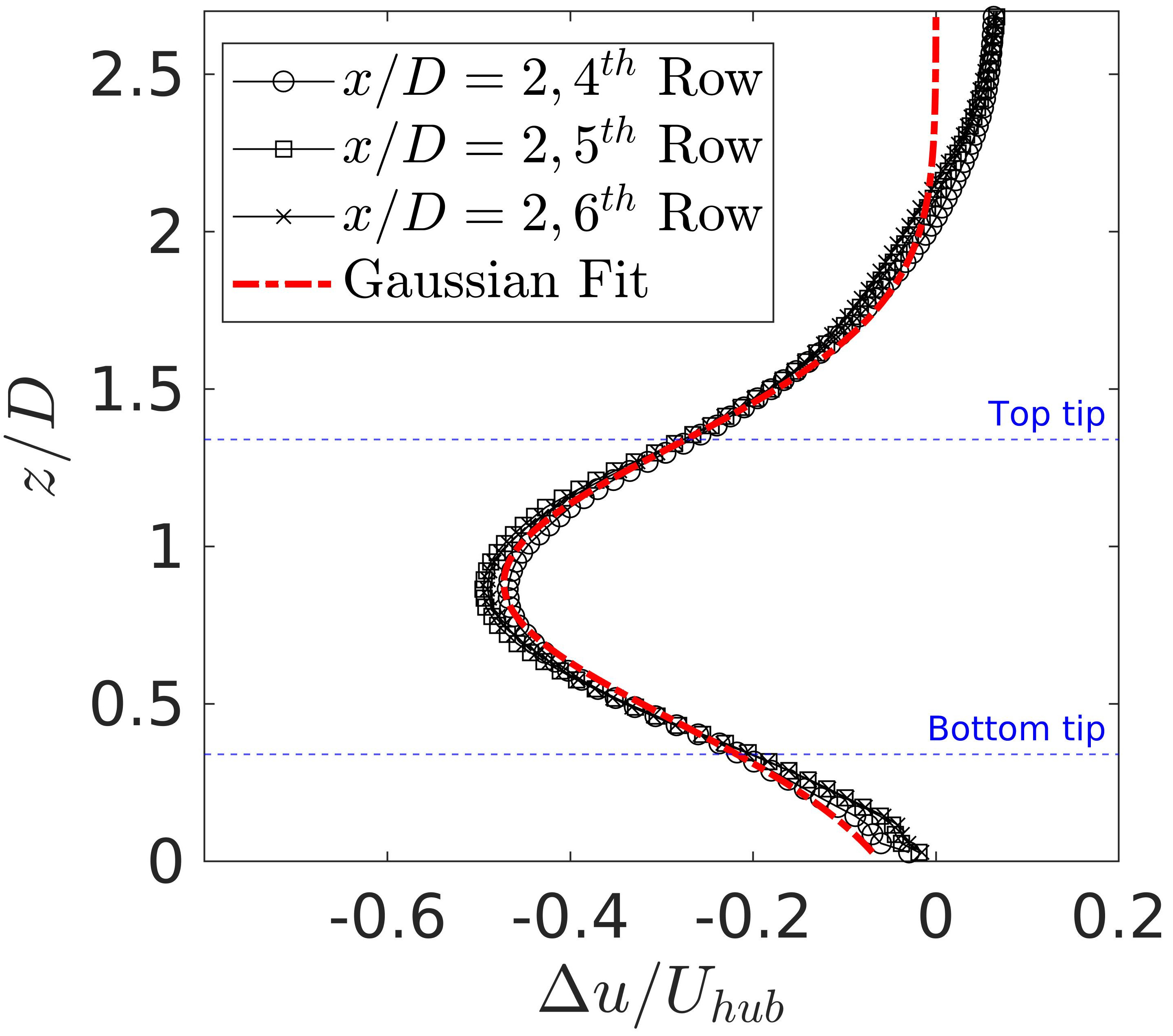}\\
		$(a)$ & $(b)$ & $(c)$
	\end{tabular}
	\caption{ $(a)$ Incoming vertical profile of the stream-wise velocity normalized with its hub-height value; $(b)$ vertical profile of the streamwise velocity in the wake behind turbines on the central column of the $10\times 3$ array at $x/D=2$ for $4^{th}$, $5^{th}$, and $6^{th}$ rows; and $(c)$ vertical profiles of the streamwise velocity deficit, {\em i.e.} the difference of the data in $(a)$ from that in ($b$). The corresponding outcome of model~B was found similar to that of model~A shown here.}
	\label{fig:flowad}
\end{figure}

\begin{figure}[ht]
	\centering
	\includegraphics[scale=0.3]{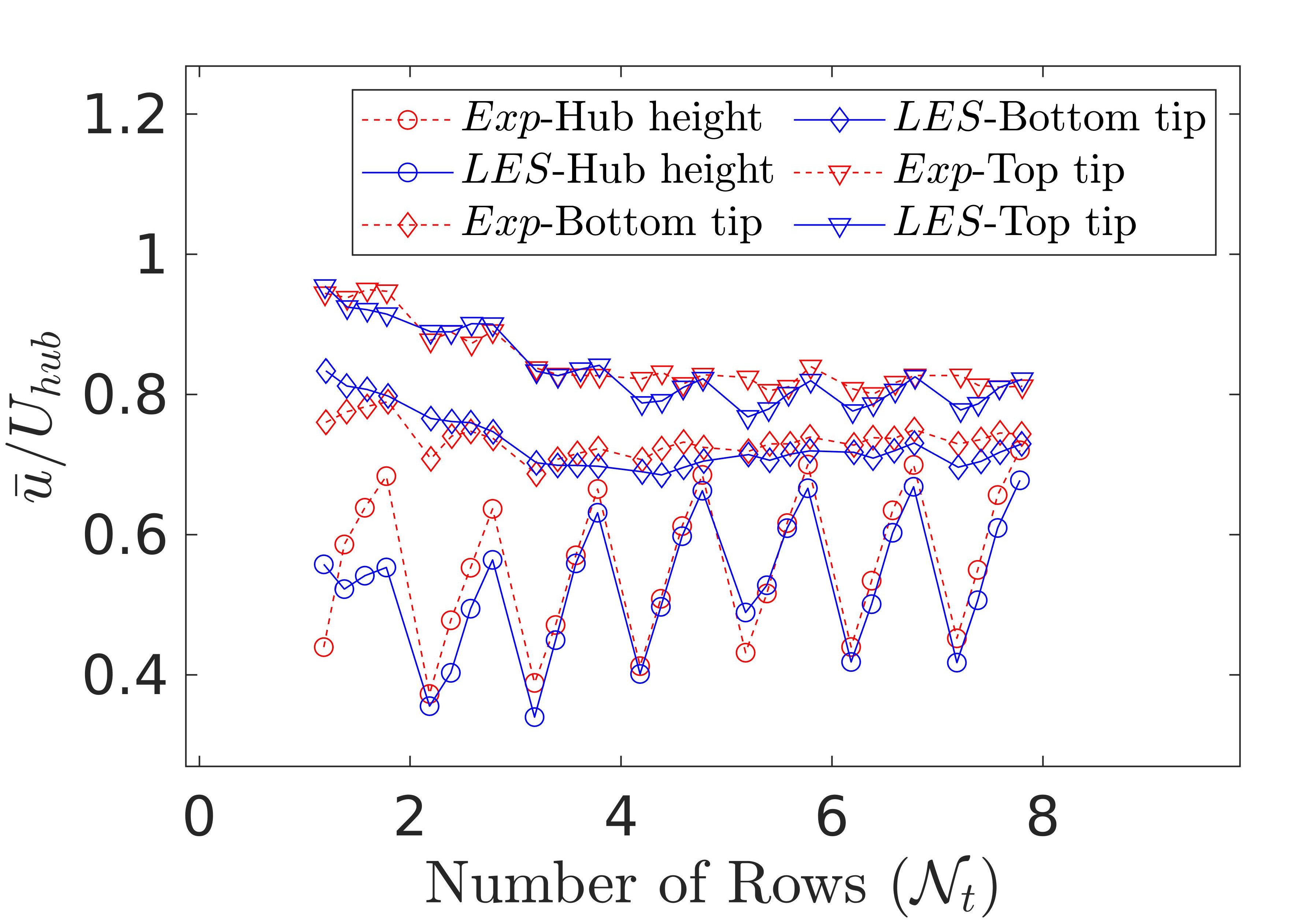}
	\caption{A comparison of the mean stream-wise velocity at top-tip, hub-height, and bottom-tip. The data correspond to velocity at four stream-wise locations: $x/D=1$, $x/D=2$, $x/D=3$, and $x/D=4$ across the central column and relative to each of the first seven rows of the wind turbine array. }
	\label{fig:stream}
\end{figure}

To illustrate the flow adjustment within wind farms, Fig \ref{fig:flowad} display the vertical profiles of the streamwise velocity and the associated velocity deficit. It is noticeable from the data that flow within the wind farm reached an equilibrium state starting from the fourth row. This result is also in a well agreement with the corresponding wind-tunnel measurements. Fig~\ref{fig:flowad}$c$ also indicates that the velocity deficit follows a perfect Gaussian distribution. Again, this indicates a further validation of the assumption made in the Gaussian actuation of the turbine in model A.

To illustrate the gradual decrease of the aerodynamic power available to each turbine, Fig \ref{fig:stream} shows the mean streamwise velocity along the top tip, hub height, and bottom tip of wind turbines. The aerodynamic power of each turbine is proportional to the third power of the velocity. Thus, comparing the wind speeds from LES with wind tunnel measurements is sufficient to validate a trend in the energy production from a numerical wind farm. Furthermore, fig \ref{fig:stream} shows that averaged streamwise velocities on the corresponding locations align with the experimental data. It is also worth mentioning that the trend of the decline in aerodynamic power depicted in Fig~\ref{fig:stream} represents that in the field measurements in the `Invenergy Vantage wind farm' in the state of Washington, USA (see Fig 5 of Ref~\cite{yang2013}).

\subsection{Turbulence in wind farms}
\label{sec:subgrid}
Atmospheric turbulence may increase or decrease the energy production of individual wind turbines. In addition, it impacts loads and fatigue of turbines and may dictate how wind farms would impact their local environment through wake effects and noise propagation~\cite{Calaf2010,roy2011simulating,churchfield2012}. In understanding turbulent flow over surface protuberances, a question often arises: how to characterize turbulent secondary motions quantifying the degree of spatial heterogeneity. Note that the scale-adaptive LES leans on a famous mathematical result. It states that the solution of the Navier-Stokes equations remains bounded as long as the enstrophy ($0.5|\bm\omega|^2$) of the flow remains bounded~\cite{Foias89}. Said another way, the production of vorticity by wind turbines (or other form of surface protuberances) would account for the stresses arising due to correlations among the spatially non-homogeneous mean horizontal and mean vertical velocities. In this section, we consider the wind-tunnel measurements of Reynolds stresses \cite{chamorro2010} for evaluating the Reynolds stresses, $\tau^R_{ij} = \widetilde{u'_iu'}_j$, resolved with proposed scale-adaptive LES method and analyze the effects of vortex stretching on $\tau^R_{ij}$. Considering turbines as probes of turbulence, we observe that a prolonged flow adjustment above the wind farm correlates with a deepening of atmospheric boundary layer (see Fig~\ref{fig:flowad}). Finally, we report relative importance of the dispersive stresses $\tau^D_{ij} = \langle u''_iu''_j\rangle$ over the Reynolds stress $\tau^R_{ij}$, where $\bar u_i(\bm x,t) = \langle\tilde{\bar u}_i\rangle(z) + u''(\bm x,t) + u'(\bm x,t)$, $\bar u_i(\bm x) = \langle\tilde{\bar u}_i\rangle(z) + u''(\bm x,t)$, and $\bar u_i(\bm x,t) = \tilde{\bar u}(\bm x) + u'(\bm x,t)$.

\begin{figure}[t]
	\centering
	\includegraphics[scale=0.22]{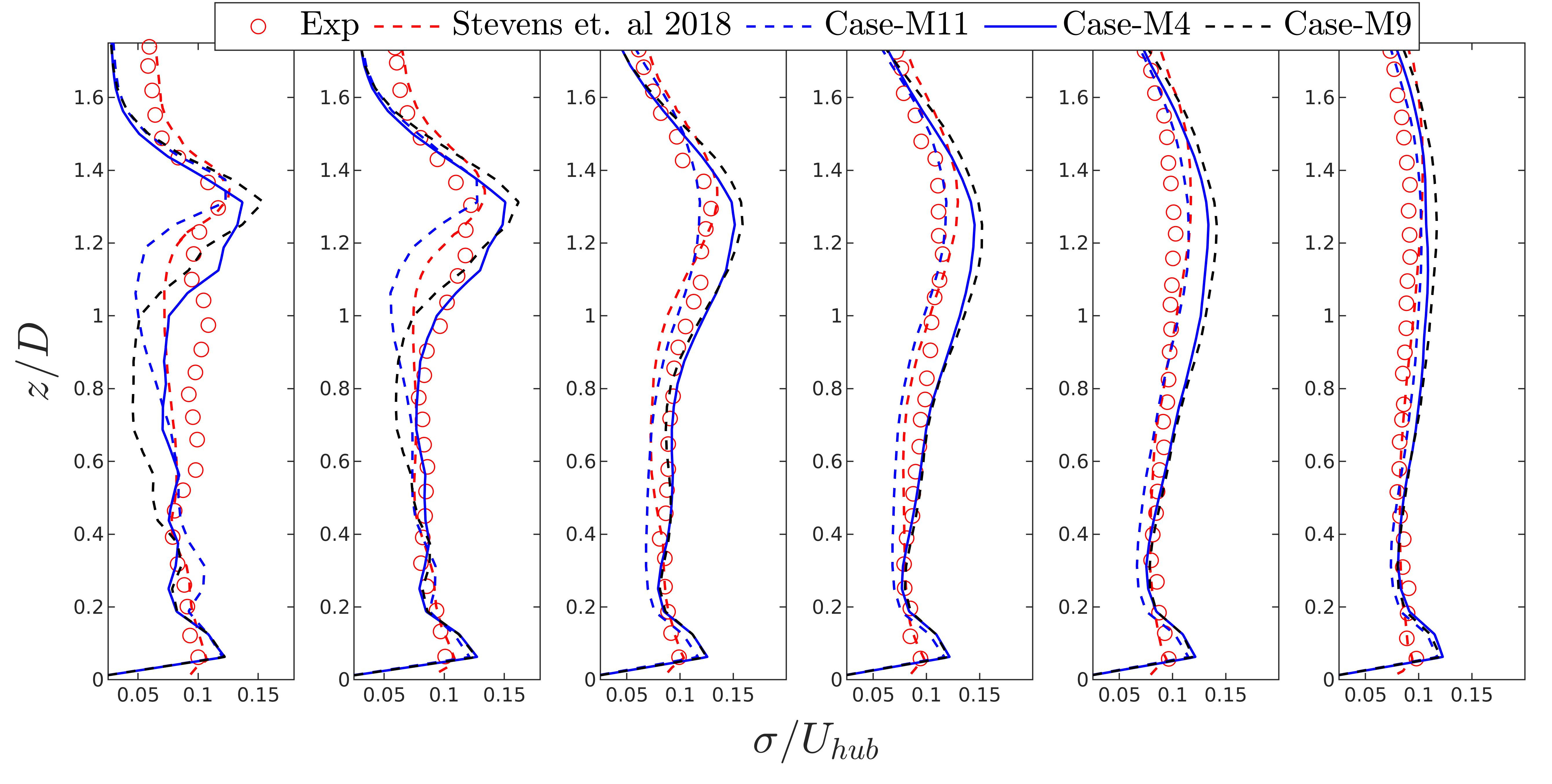}
	\caption{\label{sigmaA} A comparison of turbulence intensity ($\sigma/U_{hub}$)among wind-tunnel measurements \cite[][]{chamorro2010}, LES data \cite[][]{stevens2018}, and our present LES coupled with classical~ADM, model~A, and model~B. The data corresponds to turbulence intensity at various streamwise locations: $x/D = 2$, $x/D = 3$, $x/D = 5$, $x/D = 7$, $x/D = 10$, and $x/D=14$.}
\end{figure}
Fig \ref{sigmaA}  shows the streamwise turbulence intensity $\sigma/U_{hub}$ (where $\sigma = \sqrt{\tau^R_{11}}$) for scale-adaptive LES of a standalone wind turbine, using classical ADM, model A, and model B. We have also compared the corresponding results with wind-tunnel measurements and the LES data presented by Ref~\cite{stevens2018}. We chose ($2<x/D<7$) for the stresses shown in Fig~\ref{sigmaA} because it is a critical region for loads on a wind turbine as, in a real wind farm, a downwind turbine may be placed in this region. The maximum enhancement of turbulence intensity seems to be at the top of the wind turbine ($z/D\approx 1.3$) and is consistent with findings that appeared in the experimental investigations in the literature \cite{stevens2018,wu2011}. Numerical results obtained with three models show a similar trend of reaching the maximum value. However, they differ considerably in capturing the vertical distribution of turbulence intensity due to associated assumptions of a model. LES with model A shows a good agreement with the wind tunnel data, while classical ADM and model B slightly underpredicts the turbulence intensity in the \emph{near-wake} region ($x/D<5$). Nevertheless, all models show a similar profile in the \emph{far-wake} zone, particularly for $x/D>10$.
\begin{figure}[h]
	\centering
	\includegraphics[scale=0.25]{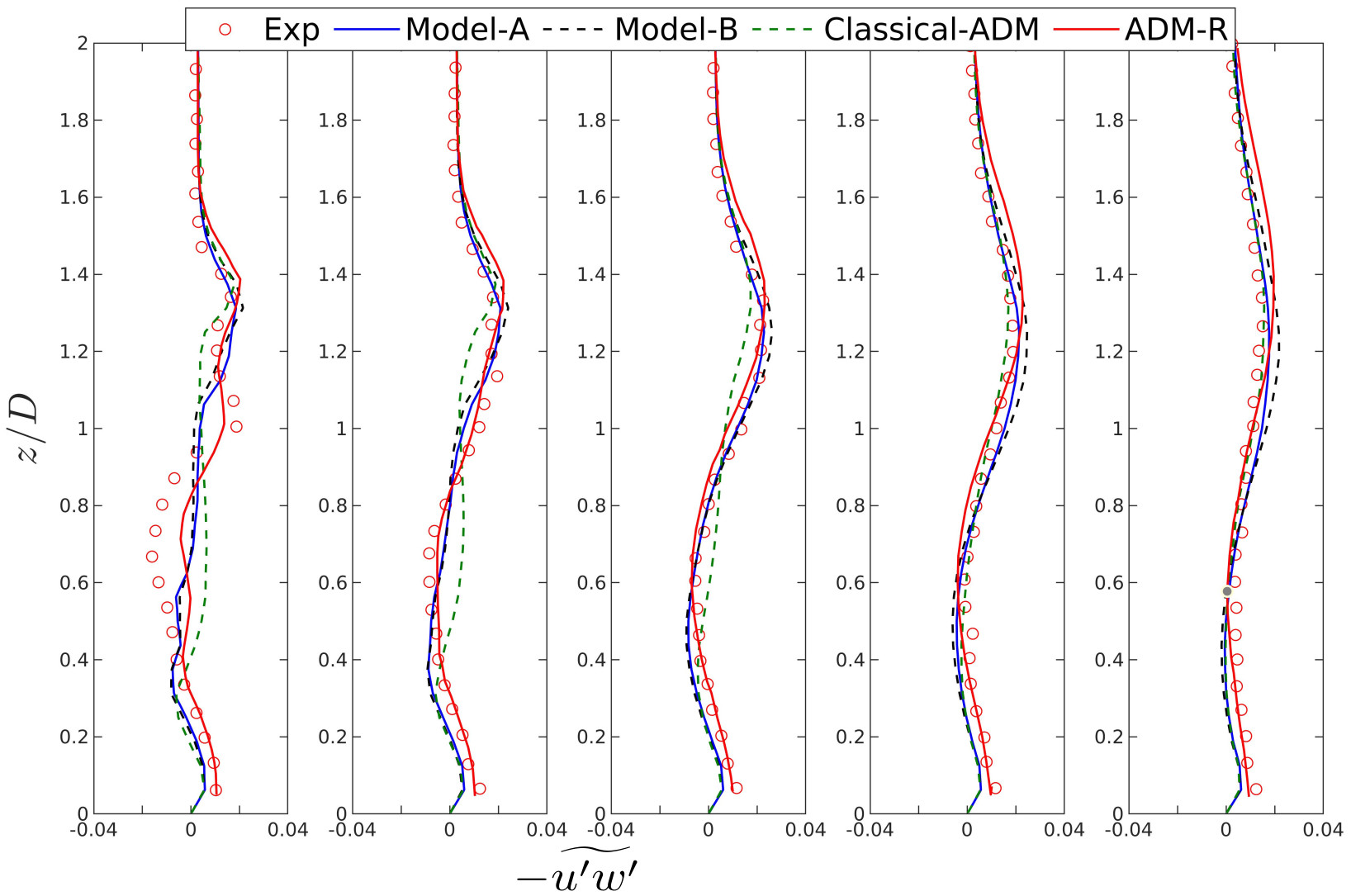}
	\caption{\label{sigma}A comparison of kinematic shear stress ($-\widetilde{u'w'}$) among wind-tunnel measurements \cite[][]{chamorro2010}, actuator disk with rotation (ADM-R) \cite[][]{wu2011}, and our present LES coupled with classical~ADM, model~A, and model~B. The data corresponds to kinematic shear stress at various streamwise locations: $x/D = 2$, $x/D = 3$, $x/D = 5$, $x/D = 7$, and $x/D = 10$.}
	\label{uw}
\end{figure}

Vertical profiles of the shear stress $-\widetilde{u'w'}$  ({\em i.e.}$-\tau^R_{13}$) appear in Fig \ref{uw}. For a pedagogical reason, we consider the corresponding result of the actuator disk model with rotation (ADM-R)~\cite{wu2011}. One observes that the shear stress has a positive value at the top tip of the wind turbine. Thus, shear production of turbulence in that region entrains high momentum air parcels from aloft and transmits toward the blades (see also \cite{Calaf2010}). On the other hand, the shear stress is negative below the hub height. Thus, horizontal fluctuations correlate highly with adverse perturbations of vertical velocity near the bottom tip of turbines, $z/D=0.3$. The outcome of this downward transmission (or ejection) of low momentum air parcels may be a potential source of ground friction velocity~$u_*$. For instance, Ref~\cite{Liu2021} observed enhanced vertical entrainment of mean energy due to an enhancement of $u_*$ in the presence of surface protrusions. 

\begin{figure}[htb!]
	\centering
	\includegraphics[scale=0.32]{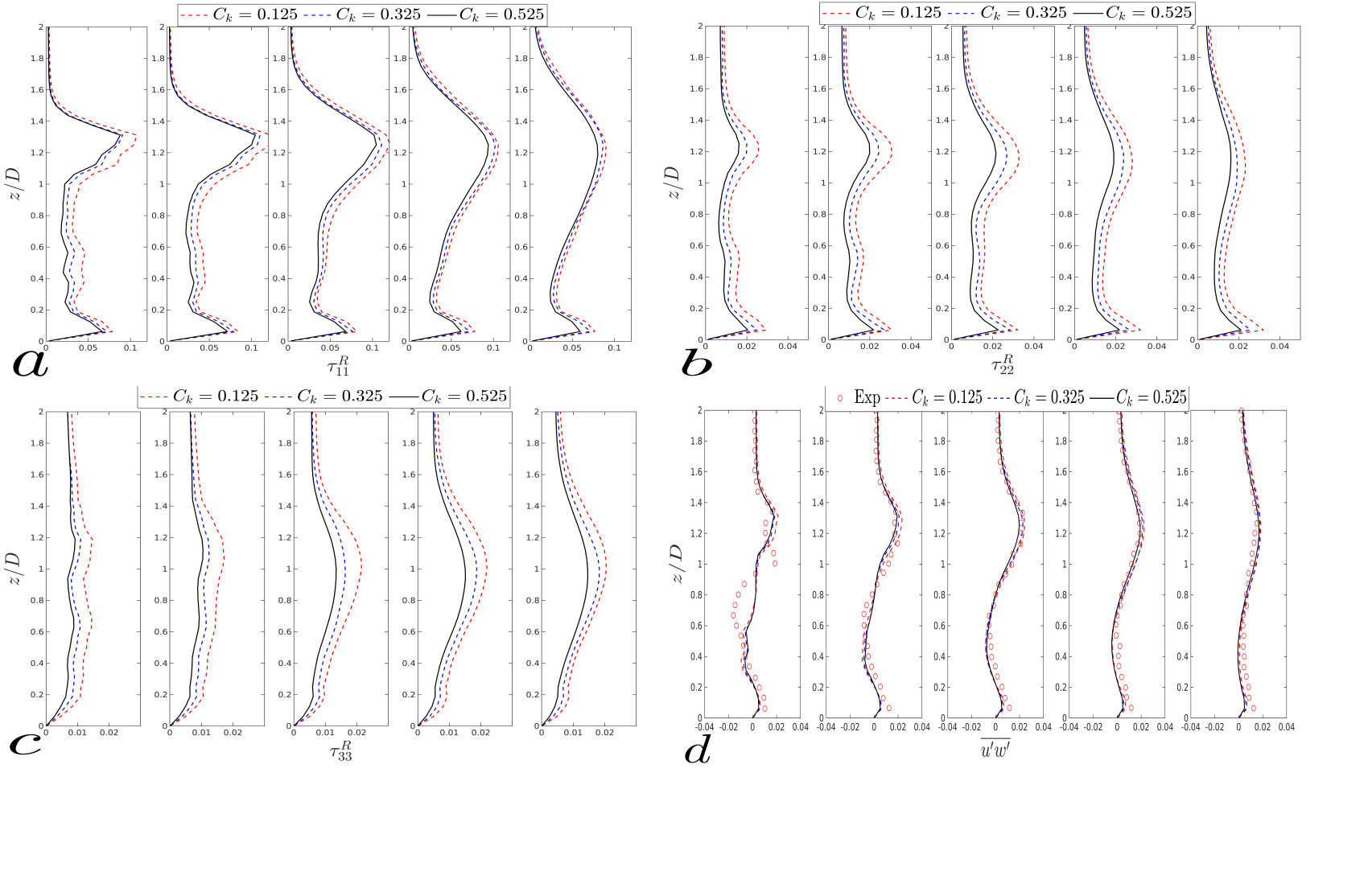}
	\caption{\label{ch1-fig4} Senstivity of model constant $c_{k}$. Data corresponds to turbulence stresses for three different model coefficient $c_{k}$ in case of scale-adaptive LES coupled with model-A at various streamwise locations: $x/D = 2$, $x/D = 3$, $x/D = 5$, $x/D = 7$ and $x/D = 10$.}
	\label{fig:totalTau}
\end{figure}

Here, we briefly evaluate the sensitivity of $c_{k}$ that appears in Eq~(\ref{eq:tau}). Fig \ref{fig:totalTau} shows the diagonal components of the resolved Reynolds stress ($\tau_{11}^{R} = -\widetilde{u'u'}$, $\tau_{22}^{R}= -\widetilde{v'v'}$, $\tau_{33}^{R} = -\widetilde{w'w'}$), and the shear stress ($\tau_{13}^{R}=-\widetilde{u'w'}$) corresponding to three values of $c_{k}$. The vertical profiles of Reynolds stresses depicted in Fig~\ref{fig:totalTau} are not drastically sensitive to a choice of $c_{k}$. This indicates that a machine learning algorithm can rigorously estimate $c_k$, if desired. However, the prediction of $c_k$ with isotropic turbulence also seems to be sufficient for the test cases considered in this article.

\begin{figure}[htb!]
	\centering
	\includegraphics[scale=0.3]{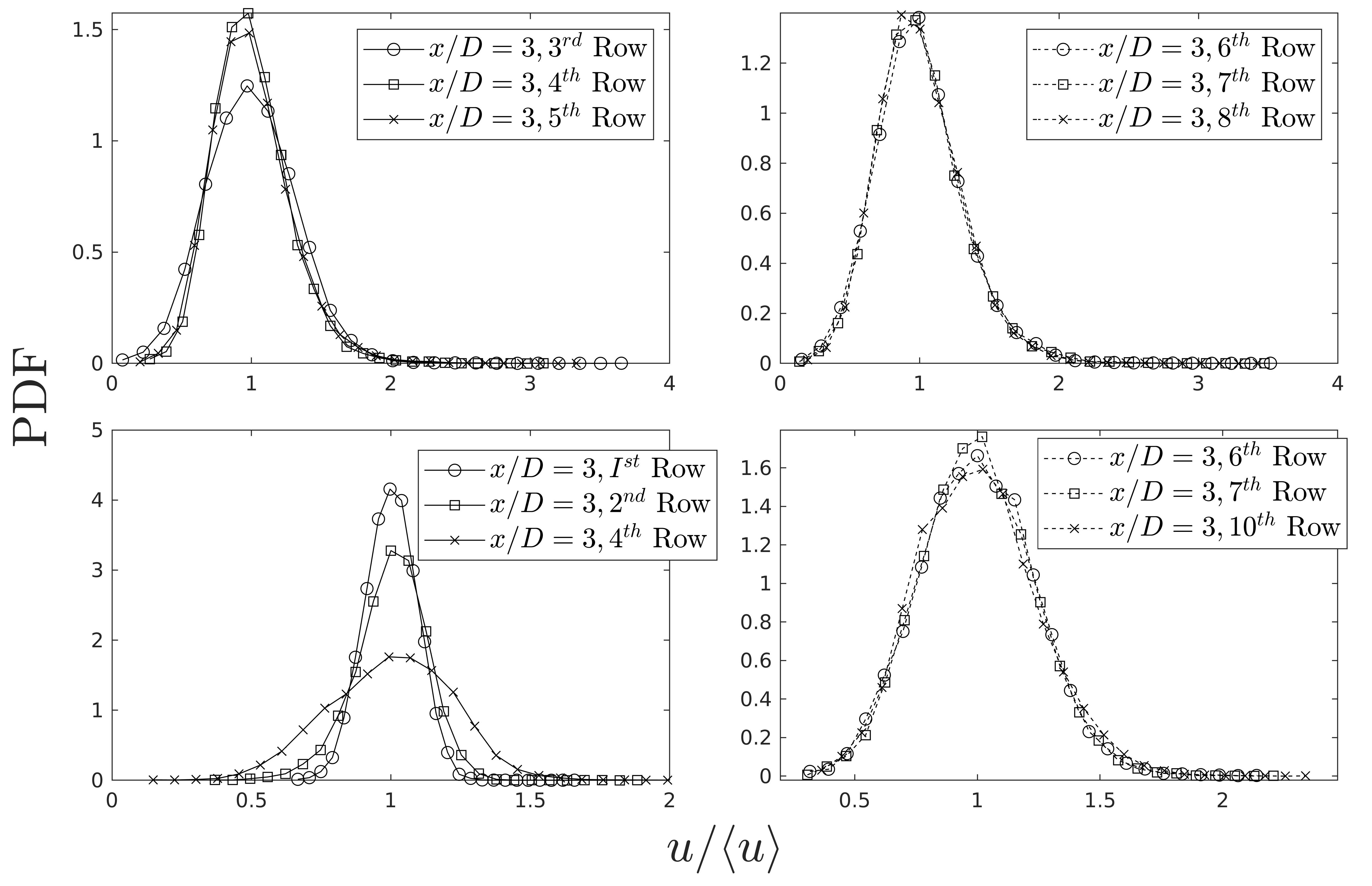}
	\caption{The top panel shows the probability density function (PDF) of instantaneously resolved velocity ($\bar u(\bm x, t)$) in region~I ($z/D <1.34$). In contrast, the bottom panel shows PDF of region~II($z/D>1.34$) in a wind farm.}
	\label{fig:pdf}
\end{figure}

According to \cite{chamorro2011}, flow in the wind farm can be split into two distinct regions based on the downstream distance needed to reach equilibrium. The region-I ($z/D<1.34$) -- just below the top tip of the wind turbine -- directly influences the wind turbines' performance. On the other hand, region-II ($z/D>1.34$) is above the wind turbine's top tip.
We collected the instantaneously resolved velocity $\bar u_1(\bm x,t)$ at $x/D=3$ and near the top tip and the bottom tip, respectively, for each wind turbine. Fig \ref{fig:pdf} shows the probability density function of normalized $\bar u_1(\bm x,t)$ in region~I and region~II. The top panel in Fig \ref{fig:pdf} reveals that flow in region~I reaches equilibrium starting from the fourth row, which indicates the quick adjustment of the flow within the wind farm. The bottom panel illustrates the probability density function of streamwise velocity at the top tip ({\em i.e.} region~II) of corresponding wind turbines. Clearly, flow adjustment above the wind farm is prolonged and does not indicate equilibrium in the first four rows and reaches equilibrium far downstream ($6^{th}-7^{th}$~row) in the wind farm, indicating that different dynamical processes are occurring between two regions, such as mixing and vertical transport. This finding agrees well with relevant experimental outcome \cite{chamorro2011}, thereby suggesting the capability of the scale-adaptive LES using model A. The point at which the flow reaches equilibrium is of significance. For example, a weather prediction model requires the input of a certain parameter treating wind farms as an elevated surface roughness. Such a similarity theory may need to consider the flow adjustments depicted in Fig~\ref{fig:pdf}.

\begin{figure}[htb!]
	\centering
	\includegraphics[scale=0.193]{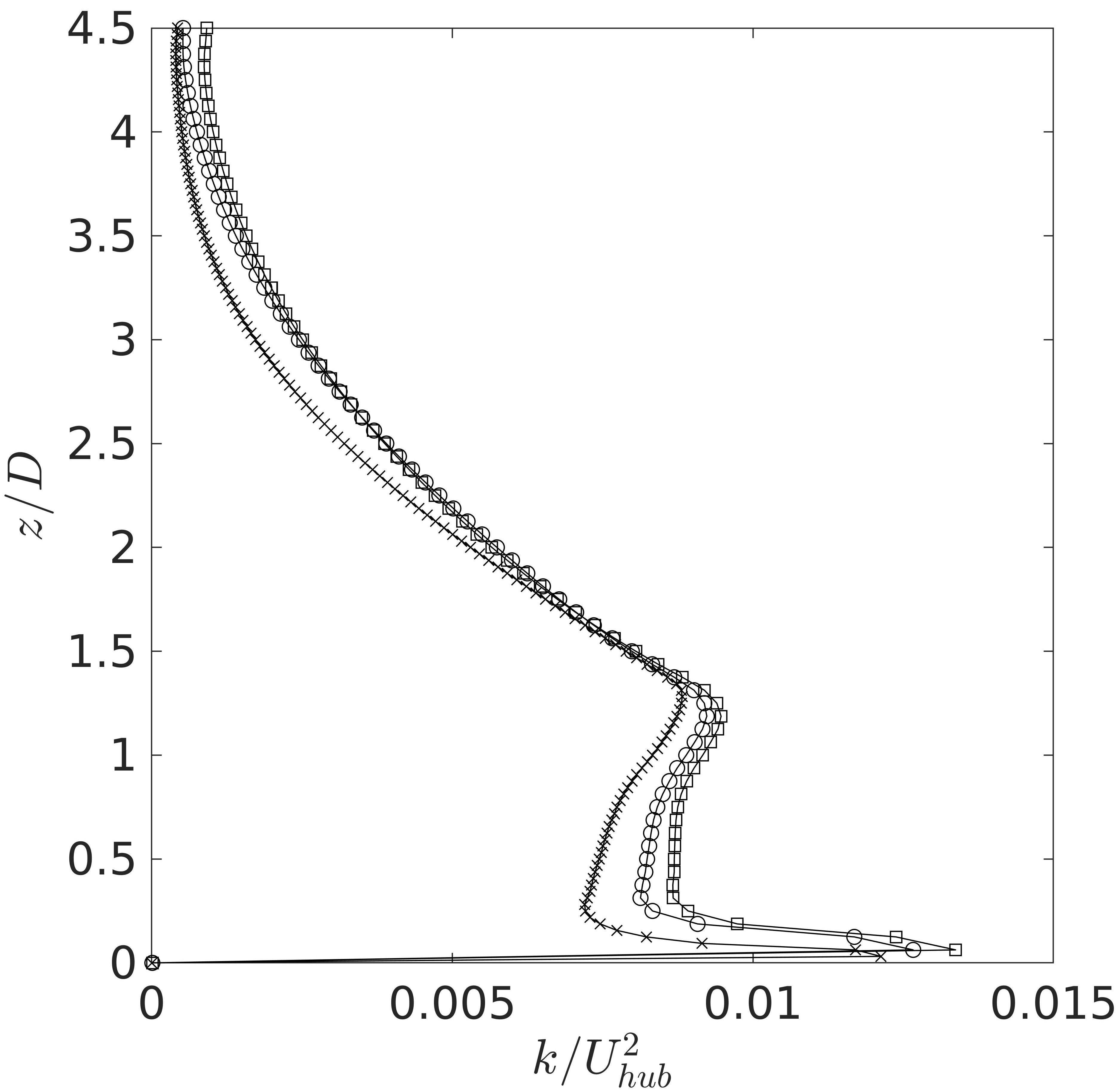}
	\includegraphics[scale=0.192]{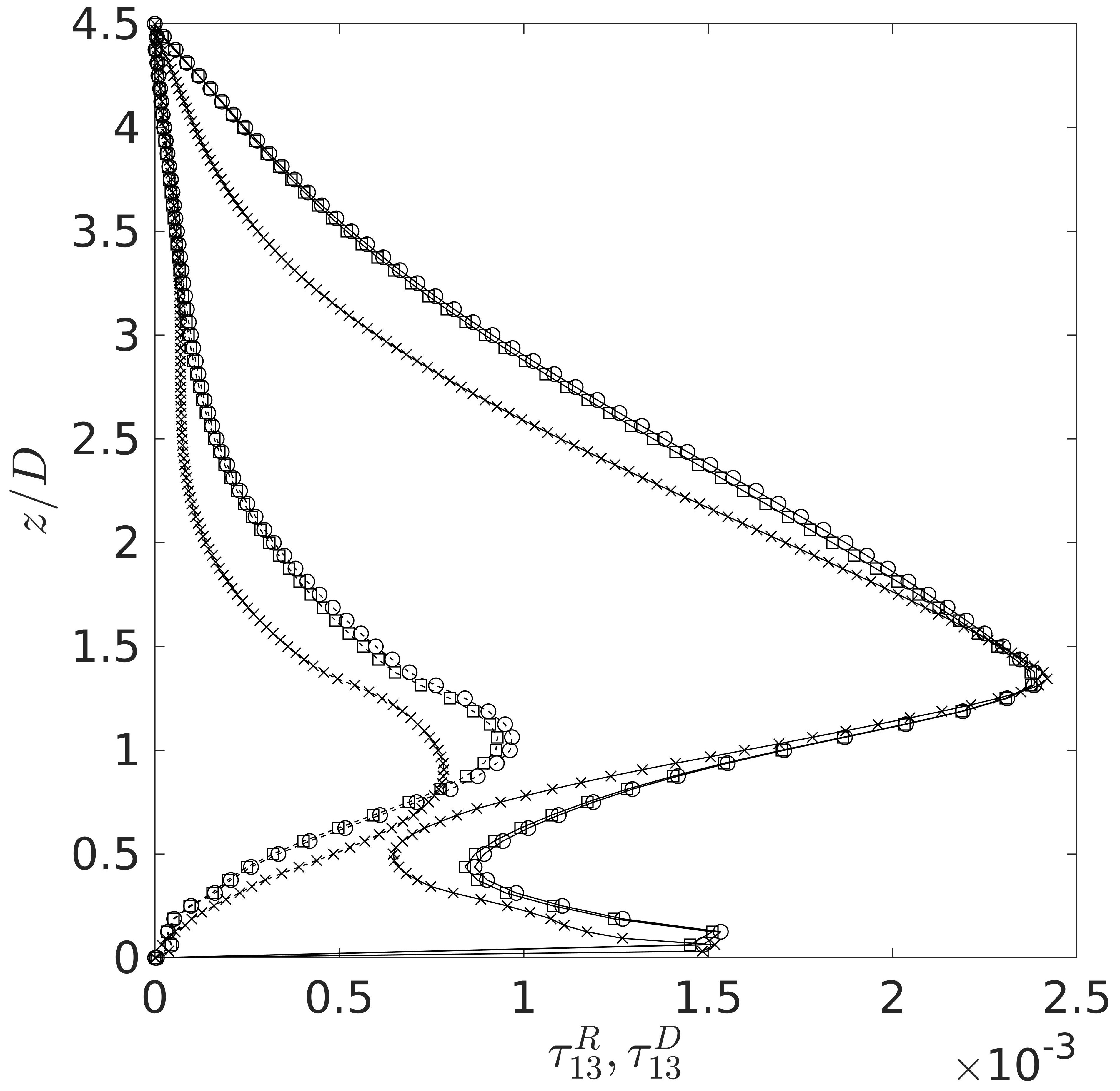}
	\caption{Vertical profiles of horizontally averaged kinetic energy ($k$), Reynolds ($\tau^R_{13}$), and dispersive ($\tau^D_{13}$) shear stress. Vertical profiles are normalized by streamwise velocity, $U_{hub}^{2}$. Here, solid lines represents Reynolds stress, whereas dashed lines represents corresponding dispersive stress and circle, square, and cross denotes for model~A, model~B, and classical~ADM, respectively.}
	\label{fig:stress}
\end{figure}

Next, we study the effect of dispersive stresses in the wind farm. The dispersive stress mainly originates from roughness elements, and their wakes \cite{jelly2019, toussaint2020}. Since the presence of a wind turbine array causes significant spatial inhomogeneity, the dispersive stress is predicted to contribute a significant portion to the total shear stress \cite{Calaf2010}. The study of spatial heterogeneity is of particular relevance because it directly impacts the wind farm's annual energy production. Therefore, it is imperative to consider the dispersive shear stress alongside the Reynolds shear stress to get a complete picture of momentum exchange in a wind farm. Plots of the spatially averaged vertical profiles of the kinetic energy $\left(k = (1/2)\text{Tr}(\tau_{ij}^{R})\right)$, Reynolds shear stress ($\tau_{13}^{R}$), and dispersive shear stress ($\tau_{13}^{D}$) for classical ADM, model A, and model B are shown in Fig \ref{fig:stress}a-b. Model A and model B predicted a maximum of $40\%$ of the dispersive stresses in the wind farm, which is consistence with the findings from other investigator~\cite{Calaf2010, poggi2004}. Dispersive stress is highly influenced by the shape, size, and how the roughness elements are distributed and thus the magnitude of dispersive stress may vary. Hence, analyzing the combined shear stresses with scale-adaptive LES may provide valuable insight into designing an efficient layout of wind farms.  

\subsection{Flow visualization}
\begin{figure}
	\centering
	\begin{tabular}{cc}
		\includegraphics[height=2.5cm, width=6cm]{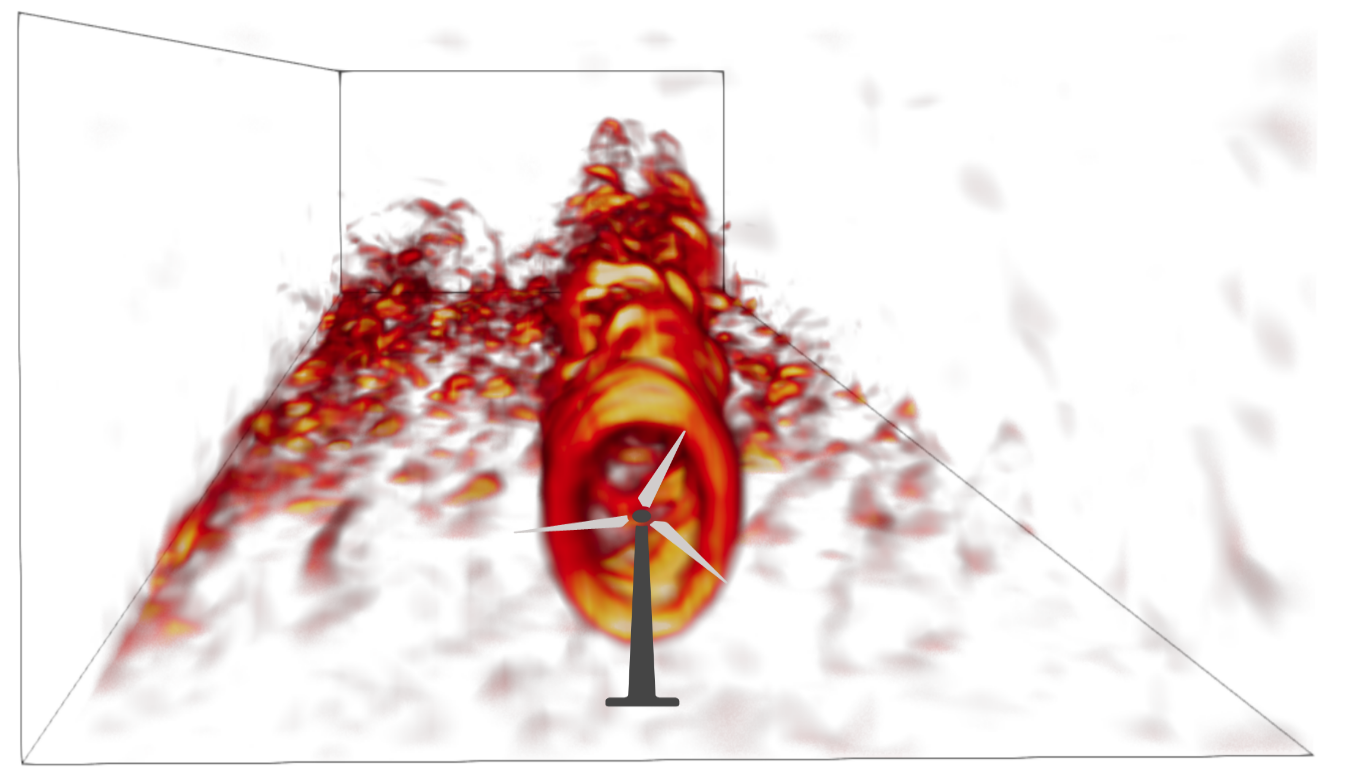}&
		\includegraphics[height=2.5cm, width=6cm]{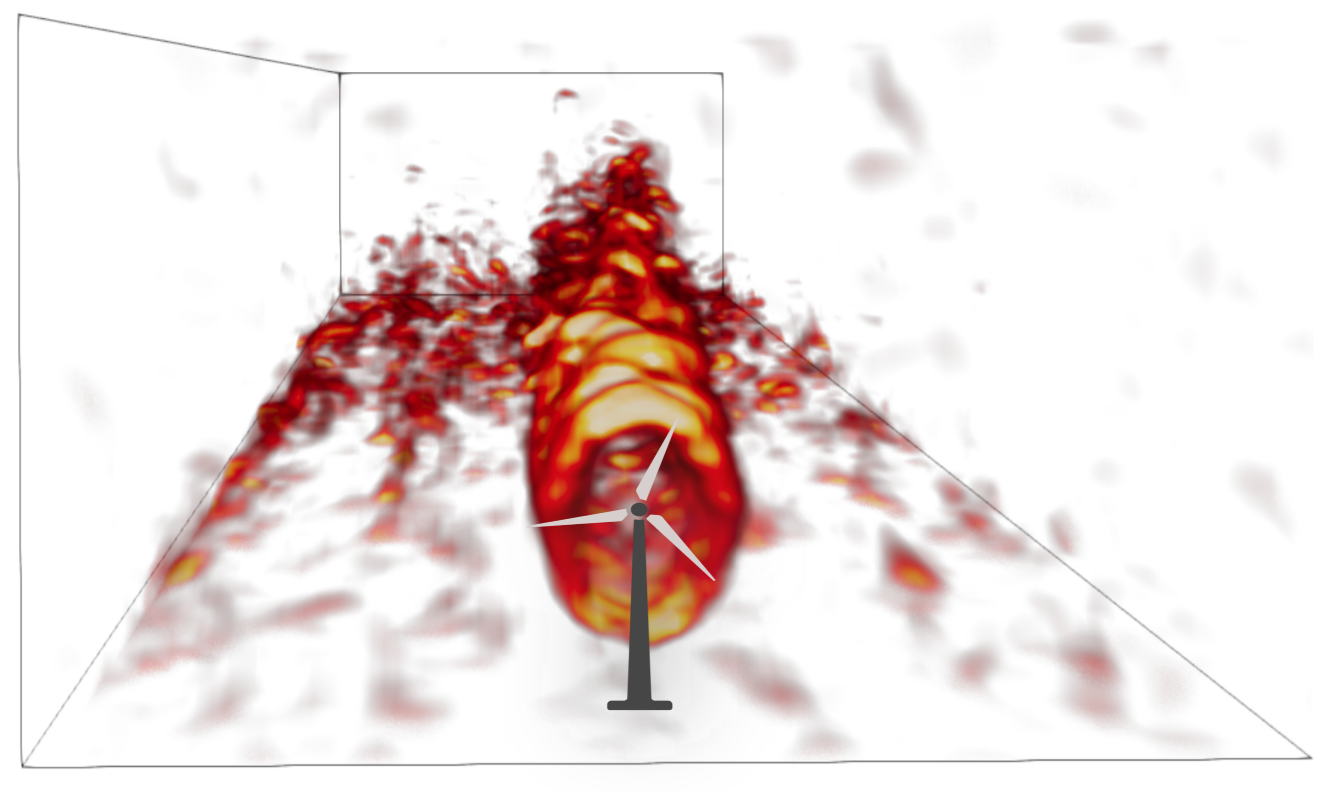}\\
		$(a)~$model~A & $(b)~$model~B
	\end{tabular}
	\caption{An isosurface plot of the magnitude of vorticity $\lvert \bm \omega \rvert$ computed from the present LES coupled with model~A and model~B.}
	\label{fig:vortswt}
\end{figure}

\begin{figure}[h!]
	\centering
	\includegraphics[scale=0.5]{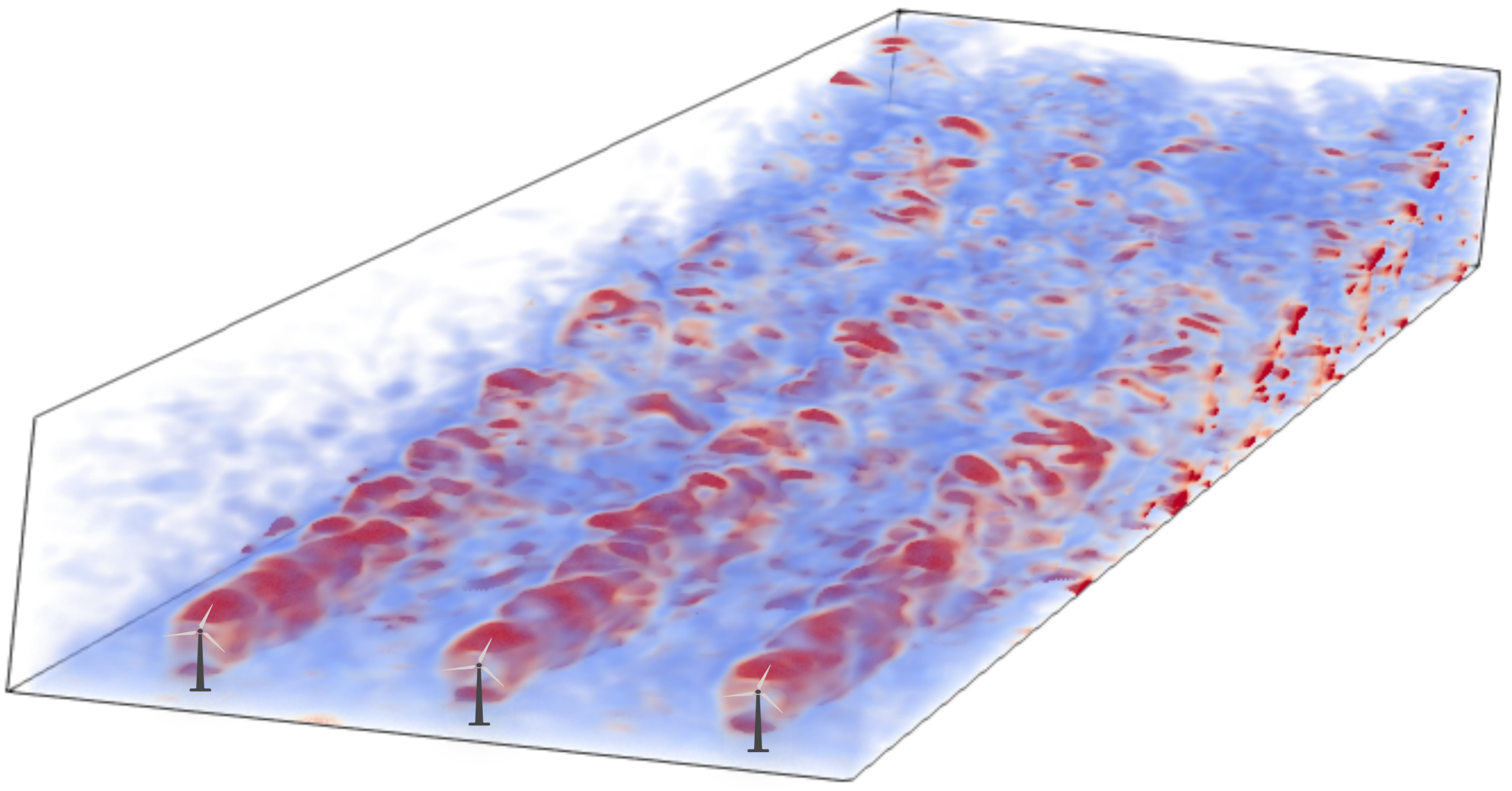}
	\caption{An isosurface plot of the magnitude of vorticity $\lvert \bm \omega \rvert$ computed from the present LES coupled with model~A.}
	\label{fig:sphereVort}
\end{figure}

\begin{figure}[h!]
	\centering
	\includegraphics[scale=0.5]{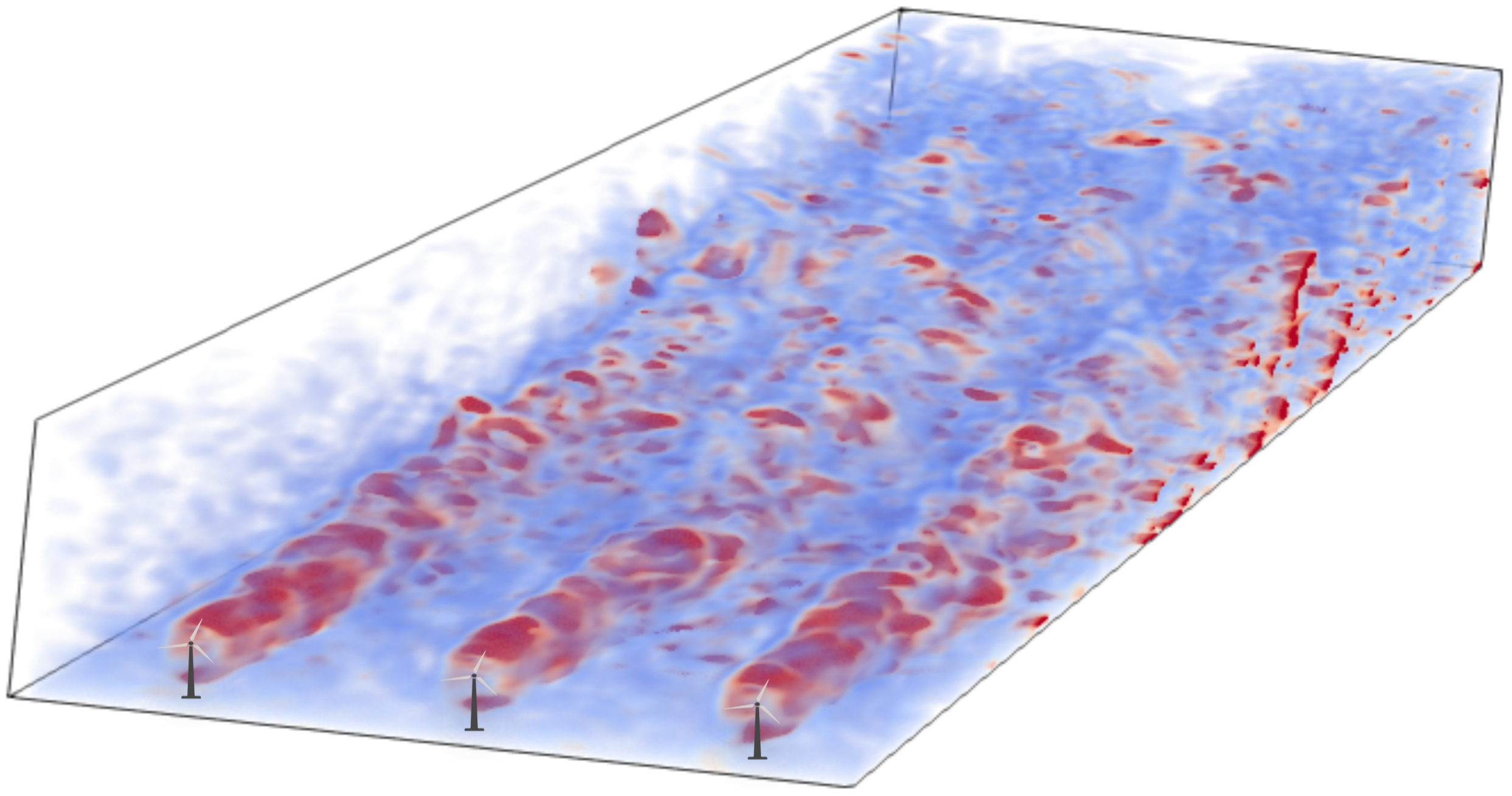}
	\caption{An isosurface plot of the magnitude of vorticity $\lvert \bm \omega \rvert$ computed from the present LES coupled with model~B}
	\label{fig:modelBvort}
\end{figure}

Fig \ref{fig:vortswt}-\ref{fig:modelBvort} shows 3D snapshots of iso-vorticity (i.e.$(1/2)\lvert \bm \omega \rvert^{2})$ for Gaussian actuator disk: `model~A' and `model~B'. It is worth mentioning that snapshots of iso-vorticity for the Gaussian ADM (Fig \ref{fig:vortswt}) closely resemble the pattern of iso-vorticity predicted with a more sophisticated actuator line model, {\em e.g.,} Fig 13 of Ref~\cite{sarlak2015}. Considering the overall dynamics of the enstrophy transport equation, 
\begin{equation}
	\label{eq:ete}
	\frac{D}{Dt}\left(\frac12|\bm\omega|^2\right) = \bm\omega^T\mathcal S\bm\omega + \frac{1}{\mathcal Re}\nabla^2\left(\frac12|\bm\omega|^2\right) - \frac{1}{\mathcal Re}||\bm\nabla\bm\omega||_1,
\end{equation}
we understand the overall rate of transfer of energy in wind turbine wakes (see Ch. 5 of Ref~\cite[][]{davidson2015turbulence} for details). It states that the growth rate of enstrophy ({\em i.e.,} the rotationality of air parcels) is strictly positive if the sum of vortex stretching  $\bm\omega^T\mathcal S\bm\omega$ and the diffusion of enstrophy $ (1/\mathcal Re)\nabla^2\left(\frac12|\bm\omega|^2\right)$ exceeds the vorticity dissipation $(1/\mathcal Re)||\bm\nabla\bm\omega||_1$. We know from mathematical analysis ({\em e.g.,} see~\cite{Foias89}) that the regularity and uniqueness of the filtered solution $\bar{\bm u}(\bm x,t)$ of the Navier-Stokes equations are guaranteed up to finite time if the enstrophy of the flow is bounded. Splitting the vorticity as $\bm\omega = \bar{\bm\omega} + \bm\omega'$, one would find two transport equations for the filtered and the fluctuating enstrophy $(1/2)|\bar{\bm\omega}|^2$ and $(1/2)|\bm\omega'|^2$, respectively. Thus, a high correlation between $\bm\omega'>0$ and $D\bm\omega'/Dt>0$ or between $\bm\omega' < 0$ and $D\bm\omega'/Dt <0$ corresponds to the growth of disturbances. This brief analysis shows that mathematical knowledge of turbulent flows helps understand the complex nature of wind turbine wakes. In particular, visualization of the enstrophy field illustrates how vortex stretching becomes important to model subgrid-scale turbulence in wind turbine wakes. 
\vspace{2cm}

\section{Conclusion and further research directions}
\label{sec:cfd}
This paper presents a detailed analysis of the scale-adaptive large-eddy simulation methodology coupled with the Gaussian actuator disk model for wind farms. To the best of our knowledge, this article contributes significantly, for the first time, toward the development of such a vortex stretching-based subgrid model. A three-dimensional Gaussian kernel is adapted to distribute the thrust force across the rotor. The accuracy and efficiency of the LES framework have been tested by comparing the model results with wind tunnel measurements and similar LES data. A detailed comparison of average streamwise velocity and Reynolds stresses reveals that the Gaussian actuator disk model performs as accurately as experiments and more sophisticated actuator line models. In addition to comparison with experimental results, we also analyzed those components that directly influence the wind farms' annual energy production (AEP). It is found that model A accurately depicts the streamwise velocity at top tip, hub height, and bottom tip. The corresponding results have an excellent agreement with the wind tunnel data. We employed probability density of instantaneously resolved velocity and observed that model A accurately predicts crucial points in interaction of atmospheric boundary layer with wind farms. Flow visualization reveals that our Gaussian actuator model predicts the wakes qualitatively similar to the actuator line model. Finding suggests that scale-adaptive large-eddy simulation, when coupled with model A, can be a promising solution for accurately assessing the annual energy production.
\cleardoublepage
\bibliography{refs}
\end{document}